\documentclass[lettersize,journal]{IEEEtran}
\usepackage{amsmath,amsfonts}
\usepackage{array}
\usepackage[caption=false,font=normalsize,labelfont=sf,textfont=sf]{subfig}
\usepackage{textcomp}
\usepackage{stfloats}
\usepackage{url}
\usepackage{verbatim}
\usepackage{graphicx}
\usepackage{cite}
\usepackage{hyperref}

\usepackage{tikz}

\usepackage{array}
\usepackage{colortbl}
\usepackage{pifont}
\usepackage{bbding}
\usepackage{makecell, booktabs, threeparttable, balance, multirow, multicol, booktabs, amsmath, amssymb, pifont, anyfontsize}

\usepackage[linesnumbered,ruled,vlined,norelsize]{algorithm2e}
\makeatletter
\newcommand{\removelatexerror}{\let\@latex@error\@gobble}
\makeatother

\usepackage{xcolor}  

\SetCommentSty{mycommfont}

\SetKwInput{KwInput}{Input}                
\SetKwInput{KwOutput}{Output}              
\SetKwInput{KwNotations}{Notations}        

\DeclareMathOperator*{\argmin}{arg\,min}

\newcommand{\ie}{\textit{i.e., }}
\newcommand{\eg}{\textit{e.g., }}

\begin{document}

\title{CoFormer: Collaborating with Heterogeneous Edge Devices for Scalable Transformer Inference}

\author{Guanyu Xu, Zhiwei Hao, 
        Li Shen,~\IEEEmembership{Member,~IEEE,} 
        Yong Luo,~\IEEEmembership{Member,~IEEE,} 
        Fuhui Sun, 
        Xiaoyan Wang, 
        Han Hu,~\IEEEmembership{Member,~IEEE,}  
        Yonggang Wen,~\IEEEmembership{Fellow,~IEEE}

\thanks{Guanyu Xu, Zhiwei Hao and Han Hu are with the School of Information and Electrionics, Beijing Institute of Technology, Beijing 100081, China. 
E-mail: \{xuguanyu, haozhw, hhu\}@bit.edu.cn.

Li Shen is with the School of Cyber Science and Technology, Shenzhen Campus of Sun Yat-sen University, Shenzhen 518107, China.
E-mail: mathshenli@gmail.com.

Yong Luo is with the School of Computer Science, National Engineering Research Center for Multimedia Software, Wuhan University, Wuhan 430072, China.
E-mail: yluo180@gmail.com.

Fuhui Sun and Xiaoyan Wang are with Information Technology Service Center of People's Court, Beijing, 100745, China. 
E-mail: sunfh6732@163.com; 428163395@139.com.

Yonggang Wen is with the College of Computing and Data Science, Nanyang Technological University, Singapore 639798.
Email: ygwen@ntu.edu.sg.
}}

\markboth{Journal of \LaTeX\ Class Files,~Vol.~14, No.~8, August~2021}%
{Xu \MakeLowercase{\textit{et al.}}: CoFormer: Collaborating with Heterogeneous Edge Devices for Scalable Transformer Inference}


\maketitle

\begin{abstract}
  The impressive performance of transformer models has sparked the deployment of intelligent applications on resource-constrained edge devices. 
  However, ensuring high-quality service for real-time edge systems is a significant challenge due to the considerable computational demands and resource requirements of these models.
  Existing strategies typically either offload transformer computations to other devices or directly deploy compressed models on individual edge devices. 
  These strategies, however, result in either considerable communication overhead or suboptimal trade-offs between accuracy and efficiency.
  To tackle these challenges, we propose a collaborative inference system for general transformer models, termed \textbf{CoFormer}. 
  The central idea behind CoFormer is to exploit the divisibility and integrability of transformer. 
  An off-the-shelf large transformer can be decomposed into multiple smaller models for distributed inference, and their intermediate results are aggregated to generate the final output. 
  We formulate an optimization problem to minimize both inference latency and accuracy degradation under heterogeneous hardware constraints. 
  DeBo algorithm is proposed to first solve the optimization problem to derive the decomposition policy, and then progressively calibrate decomposed models to restore performance.
  We demonstrate the capability to support a wide range of transformer models on heterogeneous edge devices, achieving up to 3.1$\times$ inference speedup with large transformer models. 
  Notably, CoFormer enables the efficient inference of GPT2-XL with 1.6 billion parameters on edge devices, reducing memory requirements by 76.3\%. 
  CoFormer can also reduce energy consumption by approximately 40\% while maintaining satisfactory inference performance.
  \end{abstract}


\section{Introduction}
\IEEEPARstart{W}{ith} the rapid development of 5G wireless networks and the artificial intelligence of things (AIoT), 
deploying deep neural networks (DNNs) on edge devices to enable intelligent applications has become prevalent, such as virtual/augmented reality \cite{DBLP:conf/iclr/ChenMCWP025}, autonomous vehicles \cite{DBLP:journals/pami/ChittaPJYRG23}, object tracking \cite{DBLP:conf/cvpr/MaSZ0XYY22}. 
The transformer models have emerged as the fundamental backbones in most intelligent applications for their impressive performance across diverse benchmarks \cite{swin,flan-t5}. 
However, the substantial computational demands and stringent latency requirements of transformer models pose significant challenges to deploy them on edge devices. 
For example, the GPT2-XL \cite{gpt2}, a representative transformer architecture, requires 7.8GB of memory and 3340 Giga floating point operations (GFLOPs) for inference, 
surpassing the capabilities of the typical NVIDIA Jetson Nano, which offers only 4GB of memory and a processing capability of 235.8 Giga floating point operations per second (GFLOPS).
To facilitate efficient inference, models are segmented and deployed on edge devices or cloud servers to conduct inference together, called \textit{collaborative inference} \cite{DBLP:journals/tkde/YaoZYWMZCJJSWZTKWWZY23}. 
This method helps mitigate the resource constraints and computational burdens on a single device, enabling more feasible on-device inference.

\begin{figure}
  \centering
  \includegraphics[width=0.45\textwidth]{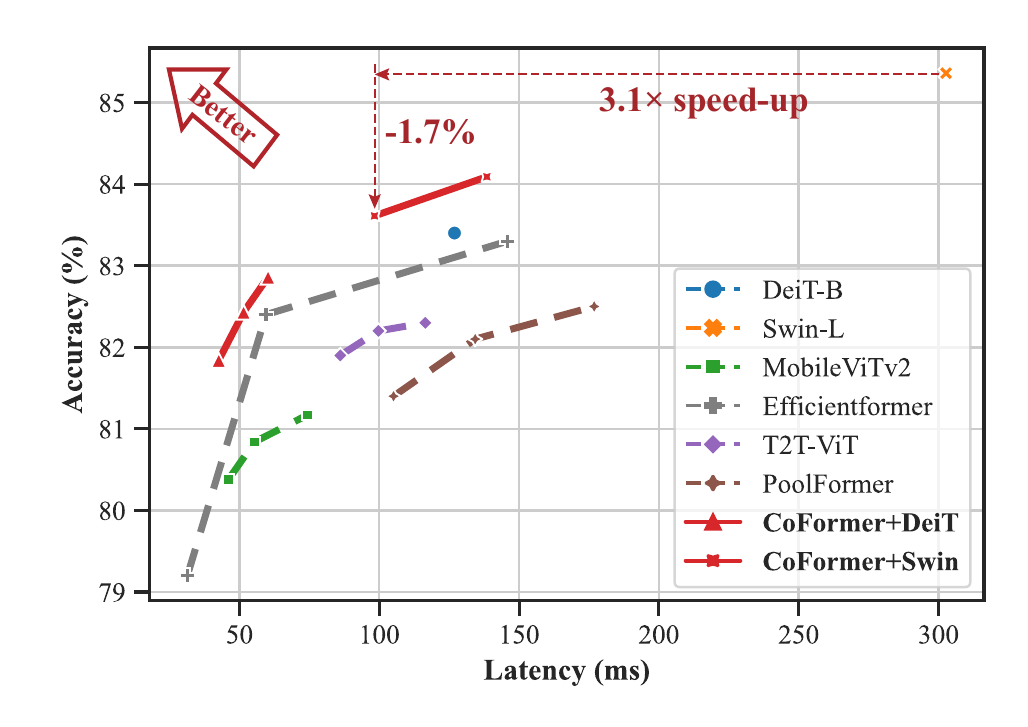}
  \vspace{-4pt}
  \caption{The comparison of classification accuracy and efficiency trade-off on ImageNet-1K. CoFormer significantly outperforms other methods. 
  Specifically, CoFormer accelerates inference speed by 3.1$\times$ compared to Swin-L \cite{swin} with only 1.7\% accuracy sacrifice. Details are shown in \S\ref{sec:performance}.  
  }
  \vspace{-4pt}
  \label{fig:trade-off}
\end{figure}

\begin{figure}
  \centering
  \includegraphics[width=0.48\textwidth]{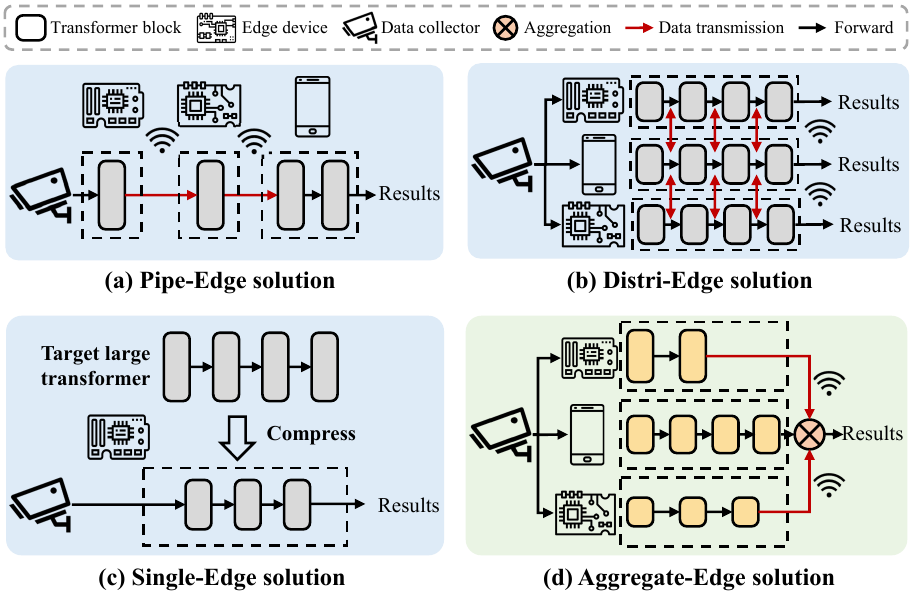}
  \vspace{-4pt}
  \caption{Comparison of collaborative inference methods.
  (a) Pipe-Edge solution: A large transformer is partitioned by layers into multiple segmented sub-models. These sub-models conduct inference sequentially. 
  (b) Distri-Edge solution: A large transformer is partitioned across layers into multiple sub-models. These sub-models conduct inference in parallel with frequent inter-communication. 
  (c) Single-Edge solution: The compressed transformer model is deployed on a single edge device for inference. 
  (d) Aggregate-Edge solution: Decomposed sub-models are deployed on edge devices and conduct inference concurrently. The intermediate results are aggregated to produce final output. 
  }
  \vspace{-4pt}
  \label{fig:existing}
\end{figure}

Existing collaborative inference methods \cite{DBLP:journals/tpds/YangZZSGL22, DBLP:journals/tmc/HaoXLHAM23, DBLP:conf/mobicom/HuangG22, DBLP:journals/ijautcomp/RenQ0JSW023} utilize a \textbf{pipe-edge solution}, depicted in Figure \ref{fig:existing} (a), which offloads tasks to cloud servers equipped with high-performance GPUs or distributes segments to various devices. 
However, this sequential execution results in high end-to-end latency. The end-to-end latency, comprising both the processing time and the communication delay among edge devices, can extend into seconds. 
A device can only commence inference after receiving outputs from preceding devices, causing significant delays. 
This delay is unacceptable for real-time interactive applications in edge systems, which typically tolerate only tens to hundreds of milliseconds of latency \cite{3gpp.22.874}. 
To minimize these delays and fully utilize computational resources across all edge devices, several researchers \cite{deepthing,coedge,galaxy} have proposed a \textbf{distri-edge solution}, illustrated in Figure \ref{fig:existing} (b).
This strategy, inspired by model parallel, involves decomposing models across layers and deployed segments on multiple devices for distributed inference. 
The decomposition scheme destroys the original structure and hinders the capability to extract information, necessitating frequent inter-device communication during inference. 
Another line of works \cite{mobilevit,mobilevitv2}, the \textbf{single-edge solution}, shown in Figure \ref{fig:existing} (c), utilize model compression \cite{deit, DBLP:conf/cvpr/YuX23,DBLP:conf/cvpr/YuCGF23} or neural architecture search (NAS) \cite{autoformerv2, ofa} techniques to obtain a compact and small model deployed on a single device. 
While this method maximizes resource utilization on a single device, striking a balance between performance and efficiency remains challenging.

Motivated by the divisibility and integrability of transformer models, 
we propose a collaborative inference system for general transformer models, termed CoFormer. It belongs to the \textbf{aggregate-edge solution} depicted in Figure \ref{fig:existing} (d), 
where an off-the-shelf large transformer model is automatically decomposed into several smaller models. 
These decomposed models can be readily deployed on edge devices for concurrent inference, and the intermediate results are transmitted only once and aggregated to produce the final output. 
The scheme significantly lessens the resource demands on any single device and reduces transmission costs in collaborative inference. 
We formulate an optimization problem to concurrently minimize inference latency and performance degradation, while accommodating the heterogeneous hardware resources. 
To address this problem, we propose the DeBo algorithm, which leverages Bayesian optimization to derive an efficient decomposition strategy.
Furthermore, DeBo incorporates a progressive knowledge distillation process to transfer knowledge from the original large transformer to the sub-models, thereby calibrating potential accuracy sacrifice caused by model decomposition.

To verify the effectiveness of our system, we conducted comprehensive experiments across three widely-adopted tasks, including image classification, object detection, and language understanding, using both vision transformer backbones (ViT \cite{vit}, DeiT \cite{deit}, and Swin \cite{swin}) and language transformer backbones (BERT \cite{bert}, GPT \cite{gpt2}, and Flan-T5 \cite{flan-t5}) on heterogeneous edge devices.
Our system supports the deployment of large transformer models on resource-constrained edge devices by significantly reducing computational burdens, thereby achieving a satisfactory balance between efficiency and performance. 
Notably, CoFormer enables the deployment of GPT2-XL, a huge transformer, on edge devices with only 4 GB memory capacity, reducing memory requirements by 76.3\%. 
Furthermore, as shown in Figure \ref{fig:trade-off}, CoFormer accelerate inference speed by 1.7$\times$ to 3.1$\times$ and reduces energy consumption by 36.3\% to 63.8\% with less than 2\% performance sacrifice compared to the large transformer. 

Our main contributions are summarized as follows:
\begin{itemize}
  \item We propose a collaborative inference system for general transformers, termed CoFormer, automatically decomposing an off-the-shelf large transformer into multiple small models deployed on heterogeneous edge devices. 
  \item We formulate an optimization problem to minimize both inference latency and accuracy degradation under the heterogeneous hardware constraints to achieve the trade-off between efficiency and performance. 
  \item We develop the DeBo algorithm to first solve the optimization problem to derive the decomposition policy, and then progressively calibrate decomposed models for performance enhancement. 
  \item We perform extensive experiments on a real-world collaborative edge computing system consisting of heterogeneous edge devices using three representative transformer backbones across three widely-used tasks, which accelerates inference speed by 2.5$\times$ and saves energy consumption by 53.2\% on average. 
\end{itemize}

\section{Rethinking Resource Utilization in Collaborative Inference}
In this section, we first define the problem encountered in collaborative inference systems and rethink resource utilization in existing collaborative inference schemes. 
Then, we are inspired by two observations to remedy the problems introduced by resource utilization. 

\subsection{Problem Definition}
\label{sec:def}
Collaborative inference \cite{DBLP:journals/tkde/YaoZYWMZCJJSWZTKWWZY23, DBLP:journals/tmc/HaoXLHAM23} involves the cooperation of multiple edge devices or cloud servers to execute model inference tasks efficiently.
In this approach, a pre-trained model is strategically decomposed, and components are distributed across various edge devices or servers.
Upon receiving an inference request, the initial segments of the model are activated to process the data. 
The resultant intermediate data is then forwarded to subsequent devices or cloud servers  for further processing. This process continues until the final result is produced.
The primary objective of collaborative inference is to minimize \textit{inference latency} and \textit{accuracy degradation} within the \textit{resource constraints} of edge devices.
Given a transformer model, the number of devices is $N$. The objective function is
\begin{equation}
    \label{eq:pro}
    \begin{aligned}
        &\min\nolimits_{\mathcal{C}} \mathcal{L}(\mathcal{C})+\delta [T_c(\mathcal{C})+T_t(\mathcal{C})],\\
        \text{s.t.}~&\text{(C0)}~r(C_n)\leq R_n, \forall n\in [1,N], C_n\in\mathcal{C},
    \end{aligned}
  \end{equation}
where $\mathcal{L}$, $T_c$ and $T_t$ represents accuracy degradation, computation latency and communication latency, respectively. 
(C0) is resource constraints of edge devices, \eg memory capacity, computational capability. 
$r(C_n)$ and $R_n$ denote the resource requirements of the sub-model and resource constraints of corresponding devices, respectively.
The decomposition policy is denoted as $\mathcal{C}=\{C_1,C_2,...,C_N\}$. Note that $\mathcal{C}=\varnothing$ represents utilizing a single edge device to execute inference. 

\subsection{Resource Utilization in Collaborative Inference}
\begin{figure}
  \centering
  \includegraphics[width=0.4\textwidth]{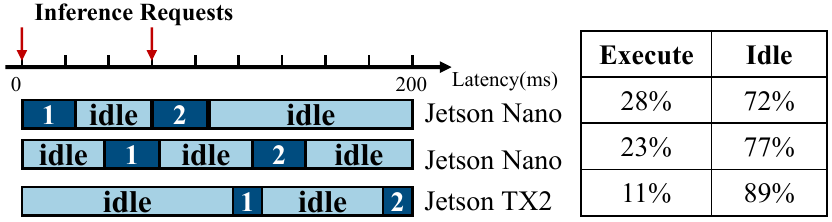}
  \vspace{-4pt}
  \caption{Latency breakdown of pipe-edge solution for DeiT-B across three edge devices. The device idleness leads to the waste of computational resources. }
  \vspace{-4pt}
  \label{fig:delay}
\end{figure}

Given the constrained hardware resources on edge devices, collaborative inference systems should aim for the trade-off between accuracy and efficiency, and maintain the quality of service (QoS) in edge computing applications. 
However, existing solutions fail to meet all goals stated in \S\ref{sec:def}, resulting in the following problems:

\textbf{Significant idle latency.}
Several methods \cite{DBLP:journals/tpds/YangZZSGL22, DBLP:journals/tmc/HaoXLHAM23, DBLP:conf/mobicom/HuangG22, DBLP:journals/ijautcomp/RenQ0JSW023} partition models by layer into multiple sequential execution blocks, termed pipe-edge solution. 
These sub-models are tailored to the resource limitations of edge devices, easing the computational load on individual devices.
However, the manner of sequential execution neglects the inherently parallel nature of neural networks, resulting in computational redundancies due to device idleness. 
We partition DeiT-B at the third and sixth layers into three sub-models and profile the inference latency on three edge devices, as shown in Figure \ref{fig:delay}.
It reveals that idle time constitutes over 70\% of the total inference process of DeiT-B on ImageNet-1K. 
The significant idle latency contributes to the overall increase in end-to-end inference latency.

\begin{figure}
    \centering
    \includegraphics[width=0.48\textwidth]{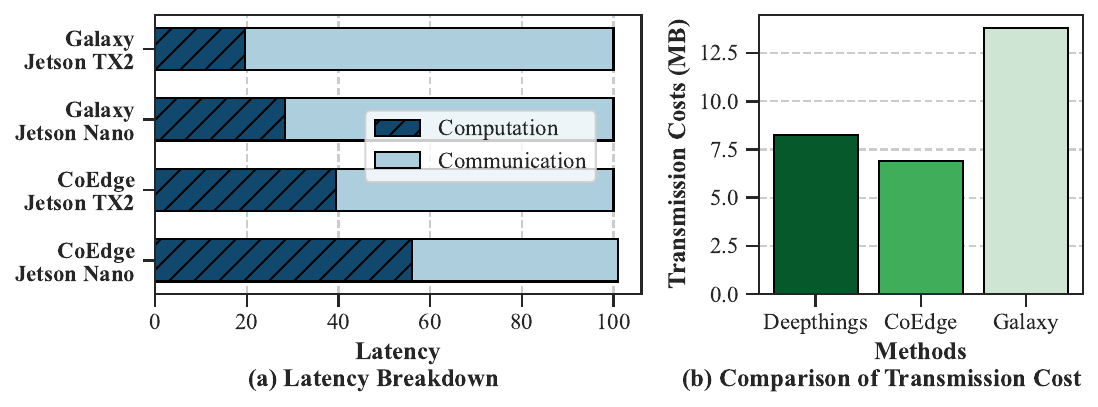}
    \vspace{-4pt}
    \caption{Inference latency breakdown (a) and comparison of transmission cost (b) for distri-edge solution using DeiT-B on ImageNet-1K.}
    \vspace{-4pt}
    \label{fig:comm}
  \end{figure}

\textbf{Substantial transmission overheads.} 
To enhance resource utilization and minimize device idleness, an alternative distri-edge solution partitions models across layers for distributed inference \cite{deepthing,coedge,galaxy, DBLP:journals/tpds/ZhouLWMW23}. 
This setup requires sub-models to frequently communicate between devices for feature interaction, 
necessitating a stable network with adequate bandwidth and resulting in significant transmission costs. 
We profile representative collaborative inference methods \cite{deepthing,coedge,galaxy} on widely-used edge devices, Jetson Nano and Jetson TX2, with a network bandwidth of 2 Mb/s and measure their transmission costs as shown in Figure \ref{fig:comm}. 
This breakdown illustrates the transmission delay constitutes more than 40\% of the total latency, and this percentage significantly increases on the Jetson TX2.
Thus, substantial network transmission delay emerges as the principal bottleneck of inference latency.

\begin{table}
\centering
\caption{Comparison of inference performance and latency for single-edge solution on ImageNet-1K.}
\label{tab:single-edge}
\resizebox{\linewidth}{!}{
    \begin{tabular}{c|c|c|c} 
        \toprule
        \textbf{Model}                      & \textbf{Device} & \textbf{Accuracy}         & \textbf{Latency}  \\ 
        \hline
        \multirow{2}{*}{EfficientFormer L7\cite{efficientformer}} & Jetson TX2      & \multirow{2}{*}{83.30\%} & 145.84\scriptsize{$\pm$4.27}ms    \\ 
        \cline{2-2}\cline{4-4}
                                            & Jetson Nano     &                           & 374.62\scriptsize{$\pm$6.41}ms     \\ 
        \hline
        \multirow{2}{*}{MobileViTv2\_200\cite{mobilevitv2}}   & Jetson TX2      & \multirow{2}{*}{81.17\%} & 74.32\scriptsize{$\pm$3.58}ms     \\ 
        \cline{2-2}\cline{4-4}
                                            & Jetson Nano     &                           & 180.83\scriptsize{$\pm$4.99}ms     \\
        \bottomrule
        \end{tabular}
}
\end{table}

\begin{figure}[t]
  \centering
  \includegraphics[width=0.45\textwidth]{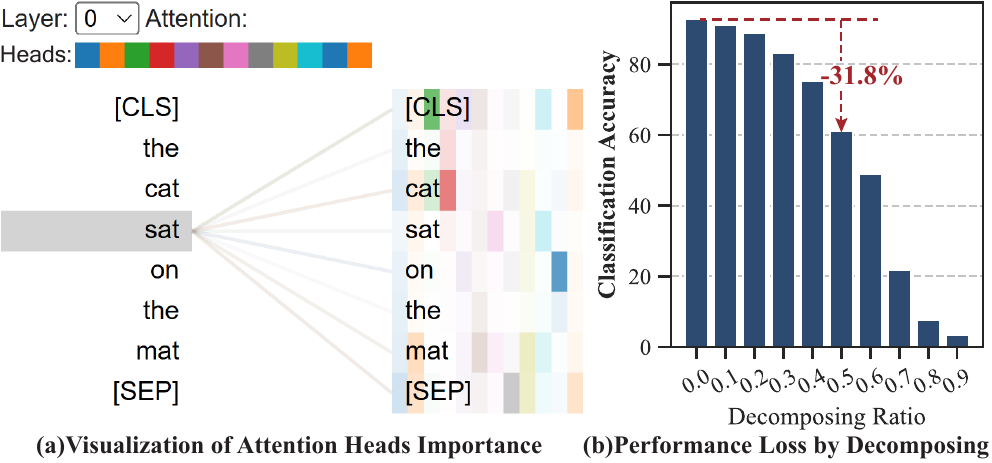}
  \vspace{-4pt}
  \caption{(a) Visualization of importance weights from attention heads in the first layer of BERT-Base. 
  The colored bands represent different attention heads, with darker colors indicating higher weight values.
  (b) Classification accuracy of BERT-Base with heads decomposing at various ratios. }
  \vspace{-4pt}
  \label{fig:divise}
\end{figure}

\textbf{Poor performance-efficiency trade-off.} 
Another line of methods such as model compression \cite{mobilevitv2,DBLP:conf/cvpr/YuX23,DBLP:conf/cvpr/YuCGF23} or NAS techniques \cite{autoformerv2, ofa} strive to optimize models to fully harness a single device's hardware resources, termed single-edge solution. 
We compare inference performance and latency of two typical methods, EfficientFormer L7 \cite{efficientformer} and MobileViTv2\_200 \cite{mobilevitv2} on the Jetson Nano and Jetson TX2 platforms using ImageNet-1K, as shown in Table \ref{tab:single-edge}. 
While EfficientFormer L7 boosts classification accuracy by 2.13\%, it also increases inference latency by 107.2\% on the Jetson Nano, a trend consistent on the Jetson TX2. 
These results highlight the challenging balance between performance and efficiency for single-edge solution.

\subsection{Insight: Divisibility and Integrability of Transformer}
\label{sec:insight}

We conduct preliminary experiments to investigate the transformer models and find the divisibility and integrability. 

\textbf{Divisibility of transformer.} 
Due to the significant computational demands of transformer models, which exceeds the capacities of individual edge devices, we propose to collaborate with multiple edge devices by leveraging the divisibility of transformer. 
Recent studies demonstrate that distinct attention heads within transformer models are capable of encoding diverse semantic information \cite{DBLP:conf/acl/VoitaTMST19, DBLP:conf/blackboxnlp/ClarkKLM19}, 
and certain neurons are critical for capturing features of particular samples \cite{DBLP:conf/cvpr/KhakzarBK00N21}.
Building on these insights, we delve into the role of different attention heads as depicted in Figure \ref{fig:divise} (a). 
Figure \ref{fig:divise} (a) reveals only a subset of attention heads are crucial for specific tokens, allowing for the feasible decomposition of heads to extract varied semantic information. 
Figure \ref{fig:divise} (b) illustrates the performance degradation resulting from this decomposition. 
A minority of heads are not essential for certain tasks, enabling their segregation with minimal performance impact. 
However,  increasing the decomposition ratio from 0.4 to 0.5 results in a 14.1\% reduction in performance, highlighting that decomposing certain target heads can lead to significant performance degradation.
Consequently, we intend to decompose a large transformer model into multiple smaller models to enable collaborative inference.

\begin{figure}
  \centering
  \includegraphics[width=0.43\textwidth]{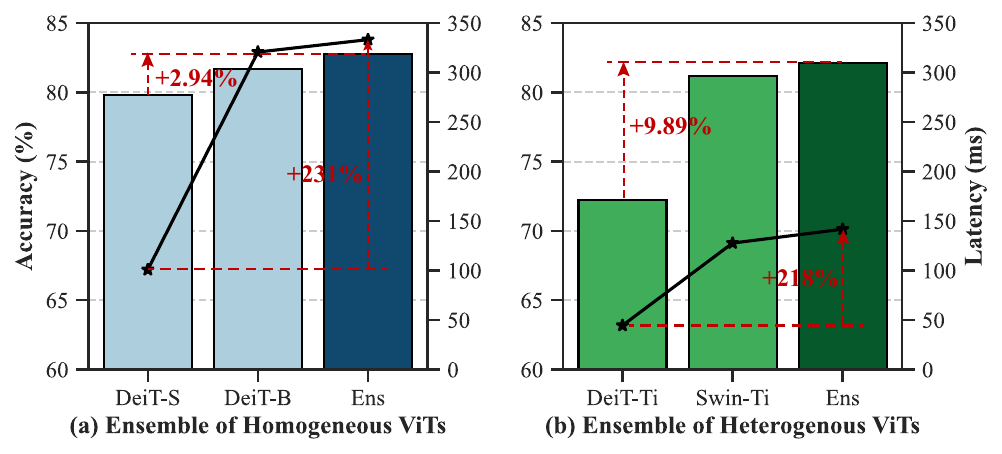}
  \vspace{-4pt}
  \caption{Ensemble performance (histogram) and latency (line chart) of homogeneous (a) and heterogeneous (b) transformer models. "Ens" represents utilize the weighted average method to aggregate these models.}
  \vspace{-4pt}
  \label{fig:integrate}
\end{figure}

\textbf{Integrability of transformer.} 
To mitigate the performance degradation resulting from decomposing large transformer models, an intuitive method is to integrate multiple decomposed models. 
Model ensemble is a strategy that combines several individual models to produce a final output \cite{DBLP:journals/air/YangLC23}, including model averaging \cite{DBLP:journals/cj/ChadhaK21}, majority voting \cite{DBLP:journals/cj/ChadhaK21}, etc. 
This technique not only boosts the overall performance but also reduces the variance across models \cite{DBLP:journals/air/YangLC23}. 
We assessed the ensemble efficacy of various transformer models through weighted averaging on the ImageNet-1K dataset, as illustrated in Figure \ref{fig:integrate} (a) and (b). 
Aggregating multiple homogeneous transformer models of varying sizes yielded a 2.94\% increase in accuracy compared to DeiT-S.
Moreover, integrating two heterogeneous transformer models led to a 9.89\% accuracy improvement over DeiT-Ti. 
However, the ensemble of multiple transformers leads to a more than 200\% increase in inference latency.
While model integration significantly boosts performance, the overall inference speed is limited by the slowest model in the ensemble.

Inspired by these two observations, 
we further propose CoFormer, a collaborative inference system by decomposing large transformer models and deploying decomposed small models on multiple heterogeneous edge devices for parallel inference. 
The intermediate results from these devices are aggregated to gain the final output. 
However, realizing CoFormer suffers from a set of challenges:
\begin{itemize}
    \item \textbf{Challenge 1: }How to decompose the large transformer model to make the most use of hardware resources among heterogeneous edge devices?
    \item \textbf{Challenge 2: }How to reduce transmission overheads during aggregating intermediate results?
    \item \textbf{Challenge 3: }How to alleviate the synchronization delay of parallel execution?
\end{itemize}

\section{CoFormer Design}
In this section, we first present an overview of CoFormer, and construct the system model to formulate inference latency and accuracy degradation minimization problem. 
Then, we illustrate the DeBo algorithm to solve the problem. 

\begin{figure}[t]
  \centering
  \includegraphics[width=0.4\textwidth]{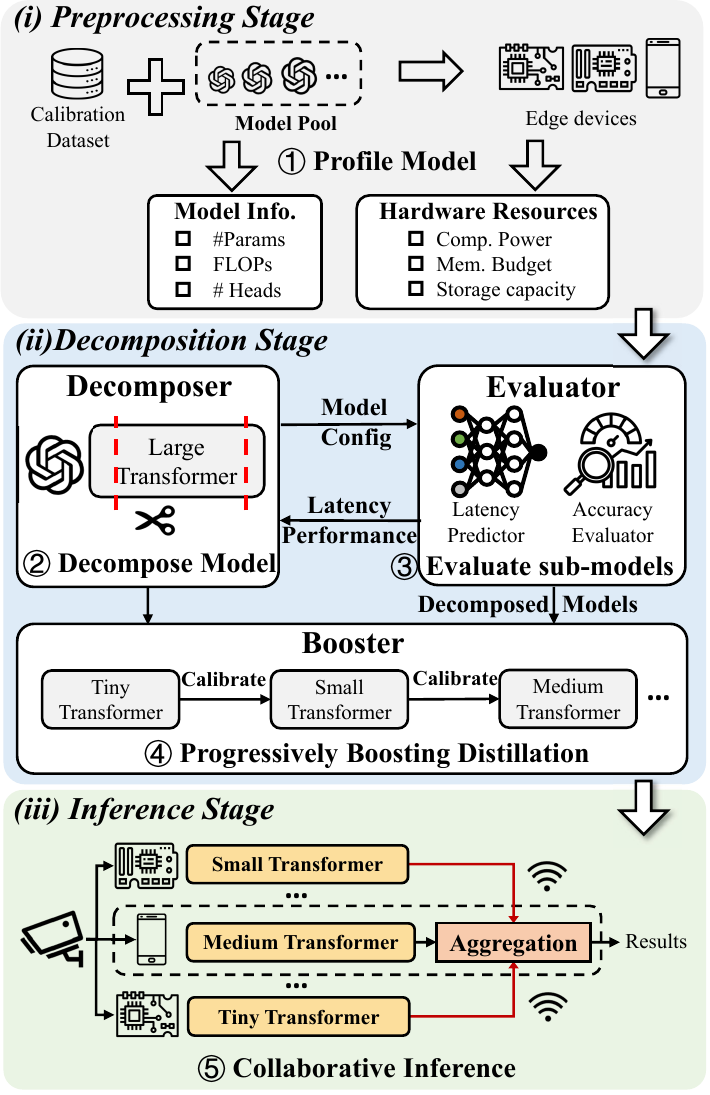}
  \vspace{-4pt}
  \caption{Overview of the proposed CoFormer.
  \textit{(i)} The transformer configurations and hardware resources are collected for the decomposition. 
  \textit{(ii)} A large transformer is first decomposed into multiple smaller models in \textit{decomposer}, which are evaluated in \textit{evaluator} until finding efficient decomposed models.
  These sub-models are progressively calibrated in \textit{booster}. 
  \textit{(iii)} These decomposed models are deployed on edge devices for collaborative inference. Intermediate results are aggregated to gain final output.
  }
  \vspace{-4pt}
  \label{fig:framework}
\end{figure}

\subsection{Design Overview}
In terms of resource-harvesting characteristics of transformer inference and existing problems of collaborative inference methods, 
we propose CoFormer, a collaborative inference system for general transformers, to achieve efficient and reliable inference on heterogeneous edge devices. 
CoFormer requires multiple heterogeneous edge devices, a central node, a target large transformer, a model pool consisting of different sizes of transformer with the same architecture as the target transformer, as well as utilizes a \textit{decomposer} to partition the transformer, an \textit{evaluator} to evaluate decomposed models and a \textit{booster} to calibrate these models. 
Figure \ref{fig:framework} depicts an overview of CoFormer which features three primary stages: 

\textit{\textbf{(i) Preprocessing Stage:} }
(\textbf{Step} \ding{192}) 
Preprocessing stage is an offline procedure that runs once before deployment. 
All edge devices execute transformers in the model pool using calibration data as input to record the runtime profile. 
The model configurations (\eg the number of model parameters, FLOPs) and hardware resources (\eg computational capabilities, memory budgets) are utilized to construct \textit{evaluator} for the decomposition stage.

\textit{\textbf{(ii) Decomposition Stage:} }
(\textbf{Step} \ding{193}) 
According to the hardware constraints of various edge devices, the large transformer is first decomposed into multiple distinct sub-models of varying sizes (\eg tiny, small, and medium transformers) by \textit{decomposer}. 
(\textbf{Step} \ding{194}) 
These sub-models are evaluated on the \textit{evaluator} to profile inference latency and accuracy. 
Evaluation results are used to update the \textit{decomposer}. The \textit{decomposer} generates new decomposition policies until finding efficient decomposed models. 
(\textbf{Step} \ding{195}) 
These models are progressively calibrated in the \textit{booster} to restore performance. 

\textit{\textbf{(iii) Inference Stage:} }
(\textbf{Step} \ding{196}) 
These calibrated models are deployed the corresponding edge devices. 
When receiving an inference task, the edge devices conduct inference with sub-models in parallel. 
Intermediate results are transmitted to central device and aggregated to produce the inference output.

\subsubsection{Inference Workflow}
The bottom of Figure \ref{fig:framework} illustrates how edge devices conduct collaborative inference tasks. 
After sub-models are calibrated and deployed on corresponding edge devices, 
inter-device communication is enabled through either wireless (\eg Wi-Fi, Zigbee, Bluetooth) or wired networks. One of edge devices is chosen as the central node. 
Upon receiving an inference task, the collected data (\eg images, videos, text) are initially transmitted across edge devices as input.
The whole inference workflow includes three phases:
\begin{itemize}
  \item \textbf{Phase 1 (Backbone Forward):} All edge devices concurrently execute the sub-models backbone for inference. 
  \item \textbf{Phase 2 (Data Transmission):} Intermediate results from other devices are transmitted to the central node. 
  \item \textbf{Phase 3 (Results Aggregation):} The central node aggregates these intermediate results to produce the inference output by utilizing \textit{aggregation} module. 
\end{itemize}

\textbf{Results aggregation.}
We present an efficient ensemble method to aggregate intermediate results from edge devices, significantly reducing transmission overhead.
The down-sampled features from the final layer of the transformer are leveraged within the \textit{aggregation} module, which substantially decreases data volume and communication latency.
Let the $i$-th device serve as the central node.
In the \textit{aggregation} module, intermediate features $\{\mathbf{X}_1,\mathbf{X}_2,...,\mathbf{X}_N\} $ from all devices are first concatenated at the central node. 
Then, an MLP block facilitates the interaction and fusion of these features through a linear transformation $\mathbf{W}\in \mathbb{R} ^{d_{agg}\times d_i }$, 
where $d_{agg}$ represents the cumulative dimension of intermediate features and $d_i$ denotes the dimension of features at the central node.
Consequently, the \textit{aggregation} module is formulated as: 
\begin{equation}
    X_{agg} = \text{Pool}(\mathbf{W} \cdot \text{Concat}\left(\mathbf{X}_1,\mathbf{X}_2,...,\mathbf{X}_N\right) + \mathbf{b} ), 
\end{equation}
where $\text{Pool}(\cdot)$ signifies an average pooling operation and $\mathbf{b}$ is the bias vector. 
The aggregated features are finally passed to the task-specific heads\footnote{The task-specific heads refer to specialized output layers in the neural network that are only responsible for making predictions for different downstream tasks. The heads are directly inherited from the large transformer model, exclude from the decomposition process.} to generate predictions. 
Our aggregation module guarantees a satisfactory inference performance with slight transmission overheads.

\subsection{Evaluator: System Modeling}
To minimize both inference latency and accuracy degradation, we pose an optimization problem focusing on these metrics to achieve efficient collaborative inference.

\subsubsection{Inference Latency Model}
A large transformer is decomposed into multiple sub-models for deployment across various edge devices. 
These devices collaboratively execute the sub-models to perform inference tasks. 
Given a system composed of $N$ edge devices, the large transformer is decomposed into $N$ sub-models under hardware constraints of edge devices. 
Each device executes its corresponding sub-model backbone for inference. We represent the set of sub-models as $\mathcal{N}=\{1,2,...,N\}$. 
The architecture configurations of sub-model $n$ are denoted as $C_n=\{l_n, d_n, h_n^{1:l_n}, D_n^{1:l_n}\}$, where $l_n$ and $d_n$ are the number of layers and embedding dimension, respectively. 
$h_n^{1:l_n}$ and $D_n^{1:l_n}$ are two vectors consisting of the number of attention heads and the MLP dimension from the first to the $l_n$-th layer, respectively. 
The set of decomposition decisions is denoted as $\mathcal{C}=\{C_1,C_2,...,C_N\}$, which indicates the large transformer is partitioned into $N$ sub-models with these specific specifications. 
When multiple edge devices execute inference tasks concurrently, their completion times can vary, requiring a delay in aggregation until the slowest device has completed its data transmission.
We can formulate the overall latency of collaborative inference as
\begin{equation}
  T=\max\nolimits_{n\in\mathcal{N}}\left(t_n^1+t_n^2\right)+t_n^3,
\end{equation}
where $t_n^1$, $t_n^2$ and $t_n^3$ represents the latency of sub-model $n$ at different phases. 

\textbf{Phase 1.} 
The inference latency $t_n^1$ for the backbone of sub-model $n$ is formulated as:
\begin{equation}
  \label{eq:t1}
  t_n^1=f\left(C_n\right)=f\left(l_n, d_n, h_n^{1:l_n}, D_n^{1:l_n}\right), \forall n\in \mathcal{N},
\end{equation}
where function $f(\cdot)$ is a latency predictor developed from actual measurement data profiled on edge devices. 
This predictor can estimate the latency of sub-models based on their configurations, as detailed in \S \ref{sec:predictor} of supplementary material. 

\textbf{Phase 2.}
The latency $t_n^2$ incurred by edge device $n$ when transmitting the intermediate feature $\mathbf{X}_n$ to the central node is calculated as follows: 
\begin{equation}
  \label{eq:t2}
  t_n^2=|\mathbf{X}_n|/r_n, \forall n\in \mathcal{N}, 
\end{equation}
where $|\mathbf{X}_n|$ signifies the size of $\mathbf{X}_n$. 
The data transmission rate between device $n$ and the central node is $r_n$. 

\textbf{Phase 3.}
The aggregated features are initially concatenated into $\mathbf{X}_{con}\in \mathbb{R} ^{S\times d_{agg}}$.
Then the features undergo a linear transformation $\mathbf{W}\in \mathbb{R} ^{d_{agg}\times d_i}$ and an average pool layer to extract the target features. 
The available computational power for the central node $i$, denoted by $g$ (in FLOPs/ms), determines the latency $t_n^3$ for results aggregation:
\begin{equation}
  \label{eq:t3}
  t_n^3=2Md_id_{agg}/g, i \in \mathcal{N}.
\end{equation}

\subsubsection{Accuracy Degradation Model}
The inference accuracy of collaborative inference primarily depends on the average accuracy of these sub-models\cite{DBLP:journals/air/YangLC23}. 
The decomposition strategy $\mathcal{C}$, determines the configurations of sub-models, which in turn affects their performance. 
Typically, generalization error is employed to evaluate the performance degradation of neural network models. 
Ideally, each model should be trained from scratch and then assessed for accuracy on a test dataset.
However, training all generated sub-models exhaustively is impractical due to significant computational overheads and time constraints.
Therefore, we adopt the average validation loss of all sub-models on the validation dataset $\mathcal{D}_{val}$, as a surrogate for generalization error to estimate accuracy degradation. 
The total accuracy degradation is defined as follows:
\begin{equation}
  \mathcal{L}_{val}=\frac{1}{N}\textstyle\sum\nolimits_{n=1}^{N}\textstyle\sum\nolimits_{x\in \mathcal{D}_{val}}\mathcal{L}_n(x,C_n) , C_n \in \mathcal{C},
\end{equation}
where $\sum_{x\in \mathcal{D}_{val}} \mathcal{L}_n$ is the validation loss of sub-model $n$.

\subsubsection{Inference Latency and Accuracy Degradation Minimization Problem}
\label{sec:problem}
Based on the above modeling, we formulate the overall optimization problem aim to minimize both inference latency and accuracy degradation in collaborative inference system.
In our scenario, each edge device receives input data at the beginning. Our goal is to find a decomposition policy $\mathcal{C}$ that optimizes both the end-to-end inference latency and accuracy degradation in collaborative inference. 
Specifically, given $N$ heterogeneous devices, a large transformer with configuration $\{L, d, h, D\}$ is decomposed into $N$ sub-models, each deployed on one of these devices.
These sub-models must meet the hardware constraints of the respective devices. Thus, the minimization problem can be formulated as follows:
\begin{equation}
  \label{eq:obj}
  \begin{aligned}
    \text{(P1)}~&\min\nolimits_{\mathcal{C}} \mathcal{L}_{val}(\mathcal{C})+\delta T(\mathcal{C})\\
    \text{s.t.}~&\text{(C1)}~l_n \leq L,~\forall n\in\mathcal N,\\
                &\text{(C2)}~\textstyle\sum\nolimits_{n=1}^{N}d_n \leq d,\\
                &\text{(C3)}~\textstyle\sum\nolimits_{n=1}^{N}h_n^k \leq h, \forall k \in [1, l_n],\\
                &\text{(C4)}~\textstyle\sum\nolimits_{n=1}^{N}D_n^k \leq D, \forall k \in [1, l_n],\\
                &\text{(C5)}~\omega(C_n)\leq \Omega_n,~\forall n\in\mathcal N, C_n \in \mathcal{C},\\
                &\text{(C6)}~\phi(C_n)\leq \Phi_n,~\forall n\in\mathcal N, C_n \in \mathcal{C},\\
  \end{aligned}
\end{equation}
where $\delta > 0$ is a balancing hyperparameter. Constraint (C1), (C2), (C3) and (C4) ensure that the number of layers, the embedding dimension, the number of attention heads and the MLP dimension of each sub-model are smaller than those of the original transformer. 
Constraint (C5) and (C6) limit the maximum computing resources and memory capacity for device $n$ to $\Omega_n$ and $\Phi_n$, respectively.

\textbf{Remark.} 
The inference latency model and accuracy degradation model are employed as the \textit{evaluator} to assess decomposition policies.
In order to solve the formulated problem, applying standard optimization methods is impractical due to two reasons:
\textit{(i)} it is a high-order optimization of $\mathcal{C}$ since we need to find all configurations of $N$ sub-models, which requires significant computational and memory costs; 
\textit{(ii)} inference latency and accuracy degradation are intractable to express in a closed-form and are not differentiable. 
To tackle these issues, we treat the objective function in equation (\ref{eq:obj}) as a black-box function and optimize it using Bayesian optimization based on the observed decomposition dynamics.

\subsection{DeBo Algorithm}
Algorithm \ref{algo:debo} presents our DeBo algorithm to optimize inference latency and accuracy degradation, 
which utilizes \textit{decomposer} to generates appropriate decomposition policies and decompose the large transformer into multiple sub-models. 
Then, \textit{booster} progressively calibrates these sub-models. 

\begin{algorithm}[!t]
  \caption{DeBo}
  \label{algo:debo}
  \DontPrintSemicolon

    \KwInput{A pre-trained large transformer model; the number of devices $N$; maximal computational cost $\Omega$; memory capacity $\Phi$;
    initial number of decomposition policies $r$; the number of search iterations $I_s$; validation dataset $\mathcal{D}_{val}$;
    weights of each training samples $W$;
    }
    \KwOutput{Calibrated sub-models;}

    \tcc{Decomposer: Bayesian Decomposition}
    
    $\{\mathcal{C}'\}_{1}^{r}\leftarrow$ randomly sample $r$ different decomposition policies satisfying $\Omega$ and $\Phi$;\\
    Utilizing $\{\mathcal{C}'\}_{1}^{r}$ to decompose the transformer model into $m$ sets of sub-models;\\
    $\{\varPsi (\mathcal{C}')\}_{1}^{r}\leftarrow$ evaluate $\{\mathcal{C}'\}_{1}^{r}$ on $\mathcal{D}_{val}$;\\
    $\text{GP}(\cdot)\leftarrow$initialize a GP prior with $\{\mathcal{C}', \varPsi (\mathcal{C}')\}_{1}^{r}$ by Equation (\ref{eq:gp});\\
      
    \For(){$i:=1$ to $I_s$}{
      $\mathcal{C}'_{i}\leftarrow$ generate the next decomposition policy by Equation (\ref{eq:acq});\\
      Utilizing $\mathcal{C}'_{i}$ to decompose the transformer model;\\
      $\{\mathcal{C}', \varPsi (\mathcal{C}')\}_{1}^{r+i}\leftarrow$evaluate $\mathcal{C}'_{i}$ on $\mathcal{D}_{val}$; \\
      Update $\text{GP}(\cdot)$ by Equation (\ref{eq:update}); \\
    }
    $\mathcal{C}^{*}\leftarrow$the decomposition policy with the minimal $\varPsi (\mathcal{C'})$ among $\{\mathcal{C}'\}_{1}^{r+i}$;\\
    Utilize $\mathcal{C}^{*}$ to decompose the transformer model into $N$ sub-models;\\
    
    \tcc{Booster: Progressively Boosting Distillation}
    Initialize the weights of training samples $W_1=\{w_i^1|w_i=1/M,i=1,...,M\}$;\\
    \For(){$j:=1$ to $N$}{
      calibrate the $j$-th sub-model using the objective function (\ref{eq:boost_loss});\\
      $W_{j+1}\leftarrow$update the weights by Equation (\ref{eq:update_weight});\\
    }
\end{algorithm}

\subsubsection{Decomposer: Bayesian Decomposition}
We utilize the \textit{decomposer} to generate decomposition policies for the large transformer as shown in Lines 1 to 11 of Algorithm \ref{algo:debo}. 
These policies are evaluated on the \textit{evaluator}. Evaluation results are utilized to update \textit{decomposer} until finding the optimal policy. 
\textit{Decomposer} decomposes the target transformer into multiple sub-models according to the final policy. 

The process of decomposing transformer models is illustrated in the \S \ref{sec:decomposer} of supplementary material due to space limitation. 
Then we introduce how to gain the optimized decomposition policy. 

As analyzed in \S\ref{sec:problem}, we treat the objective function in the equation (\ref{eq:obj}) as a black-box-function $\varPsi (\mathcal{C})=\mathcal{L}(\mathcal{C})+\delta T(\mathcal{C})$ and employ BO technique to optimize. 
It applies exploration and exploitation to the objective function by sequentially and actively querying the function values of some input instances. 
In this work, we utilize Gaussian process (GP) as a \textit{prior model} to fit the black-box objective function $\varPsi (\mathcal{C})$ in a closed form. 
It updates the distribution of $\varPsi (\mathcal{C})$ to gain \textit{posterior prediction} by using its likelihood on newly evaluated $(\mathcal{C}, y=\varPsi (\mathcal{C})+\epsilon)$ pairs, 
where $y$ is a noisy observation of $\varPsi (\mathcal{C})$ and is the sum of inference latency and accuracy degradation in our case. 
Then it determines the next decomposition policy $\mathcal{C}$ by minimizing an \textit{acquisition function}, 
which is computed from the updated posterior. The \textit{acquisition function} performs a trade-off between exploration and exploitation in evaluating the candidates of decomposition policy. 
The above process is repeated until achieving a precise posterior predictive distribution of $\varPsi (\mathcal{C})$.

\textbf{Prior model.} GP is utilized as the prior to model the objective function. 
A GP prior is specified by its mean function $\mu(\cdot)=\mathbb{E} [\cdot]$ and covariance function $K(\cdot,\cdot)$. 
The prior of objective function is defined as follows:
\begin{equation}
  \label{eq:gp}
  \begin{aligned}
    &\varPsi (\mathcal{C}) \thicksim \text{GP}(\mu(\cdot), K(\cdot,\cdot)), \\
    &K(\mathcal{C}_i,\mathcal{C}_j)=\frac{\varTheta^{\nu} K_{\nu}\left(\varTheta\right)}{\Gamma(\nu)2^{\nu-1}}, \varTheta=\sqrt{2\nu ||\mathcal{C}_i-\mathcal{C}_j||_2}/l.
  \end{aligned}
\end{equation}
We set $\mu(\cdot)=0$ and $K(\cdot,\cdot)$ to be the Matern kernel with smoothness factor $\nu =1.5$ and length scale $l=1$. 
$K_{\nu}(\cdot)$ and $\Gamma(\cdot)$ are a modified Bessel and gamma function, respectively.

\textbf{Posterior prediction.}
We sample $r$ different decomposition policy $\mathcal{C'}$ satisfying the constraints $\varOmega =\{\Omega_1,...,\Omega_N\}$ and $\varPhi=\{\Phi_1,...,\Phi_N\}$ of all devices, 
and obtain $r$ sets of decomposed sub-models. 
For notation simplicity, vectors composed of $\{\mathcal{C}'\}_{i=1}^{r}$ and $\{\varPsi (\mathcal{C}')\}_{i=1}^{r}$ are denoted as $\mathcal{C}'_{1:r}$ and $\varPsi (\mathcal{C}'_{1:r})$, respectively. 
We evaluate the objective function denoted by $\hat{y}_i$ as a noisy observation of $\varPsi (\mathcal{C}')$, \ie $\hat{y}_i=\varPsi (\mathcal{C}')+\epsilon$ where Gaussian white noise $\epsilon \thicksim \mathbf{N}(0, \sigma^2)$. 
Given the noisy observations $\hat{y}_{1:r}$, the GP posterior can be updated as follows:
\begin{equation}
  \label{eq:update}
  \varPsi (\mathcal{C}'_{1:r})|\mathcal{C}'_{1:r}, \hat{y}_{1:r} \thicksim \mathbf{N}(\hat{y}_{1:r}, \mathbf{K}+\sigma^2\mathbf{I}).
\end{equation}
Given a new decomposition policy $\mathcal{C}$ and $\mathbf{k}_i=K(\mathcal{C},\mathcal{C}'_i)$, the GP posterior can be used to predict the distribution of $\varPsi (\mathcal{C})$: 
\begin{equation}
  \begin{aligned}
    \varPsi (\mathcal{C})|\mathcal{C}'_{1:r},\hat{y}_{1:r} &\thicksim \mathbf{N} (\mu(\mathcal{C}), \sigma^2(\mathcal{C})),\\
    \mu(\mathcal{C}) & \triangleq \mathbf{k}(\mathbf{K}+\sigma^2\mathbf{I})^{-1}\hat{y}_{1:r},\\
    \sigma^2(\mathcal{C}) & \triangleq K(\mathcal{C}, \mathcal{C})-\mathbf{k}^T (\mathbf{K}+\sigma^2\mathbf{I})^{-1}\mathbf{k}.
  \end{aligned}
\end{equation}

\textbf{Acquisition function.}
Given the posterior predictive distribution of $\varPsi(\mathcal{C})$, BO finds the next decomposition policy $\mathcal{C}'_{i+1}$ to evaluate based on an acquisition function $u_i(\mathcal{C})$ defined by the posterior mean $\mu_i(\mathcal{C})$ and standard deviation $\sigma_i(\mathcal{C})$. 
In our algorithm, expected improvement (EI) is utilized as acquisition function. 
For problem (P1), we can determine the next decomposition policy $\mathcal{C}'_{i+1}$ by minimizing EI:
\begin{equation}
  \label{eq:acq}
  \begin{aligned}
    \mathcal{C}'_{i+1} &= \argmin_{\mathcal{C}} u_i(\mathcal{C}), \\
    u_i(\mathcal{C}) &\triangleq (\varPsi^{*}-\mu)Z ((\varPsi^{*}-\mu)/\sigma) + \sigma H ((\varPsi^{*}-\mu)/\sigma),
  \end{aligned}
\end{equation}
where $Z$ and $H$ represent the probability density function and cumulative distribution function of the standard normal distribution, respectively. 
$\varPsi^{*}$ is the best policy in $\mathcal{C}'_{1:r}$.

\textbf{Analysis of search space.}
Consider a large transformer model with $L$ layers, $h$ attention heads, embedding dimension $d$, and MLP hidden dimension $D$. The configuration of a sub-model deployed on device $n$ is defined as $C_n = {l_n, d_n, h_n^{1:l_n}, D_n^{1:l_n}}$, where $l_n$ and $d_n$ denote the number of layers and embedding dimension, respectively. 
$h_n^{1:l_n}$ and $D_n^{1:l_n}$ are two vectors consisting of the number of attention heads and the MLP dimension from the first to the $l_n$-th layer, respectively. 
This configuration must satisfy the hardware constraints of device $n$, including memory $\phi(C_n) \leq \Phi_n$ and computation $\omega(C_n) \leq \Omega_n$. Accordingly, the search space can be characterized as $l_n^3 d_n h_n^{1:l_n} D_n^{1:l_n}$, constrained by both memory and computational budgets.
The decomposition is an offline procedure that runs once before deployment. 

\subsubsection{Booster: Progressively Boosting Distillation}
To mitigate the performance degradation resulting from decomposition, we introduce the \textit{booster}, employing a progressively boosting distillation method to improve performance as shown in Lines 12 to 15 of Algorithm \ref{algo:debo}. 

The booster consists of a large transformer, $N$ decomposed sub-models and a training dataset.
We denote the training dataset as $\mathcal{D}_{train}=\{(x_i,y_i), i\in[1,M]\}$, where $M$ and $R$ are the number of samples and the total number of sample classes, respectively.
For the $n$-th iteration, the weights of all training samples are $W_n=\{w_i^n|i=1,...,M\}$.
The progressively boosting distillation calibrates each sub-model sequentially, comprising two primary stages per iteration: weight update and model distillation. 

\textbf{Weight update.}
In each iteration, the weights of training samples are adjusted based on the prediction errors of the last calibrated sub-model. 
Note that in the first iteration, sample weights are uniformly initialized as $W_1=\{w_i^1|w_i=1/M,i=1,...,M\}$.
For the $n$-th iteration, the sample weights $W_n$ are updated as follows:
\begin{equation}
  \label{eq:update_weight}
  w_i^{n}=w_i^{n-1}\cdot \text{exp}\left[(1/M-1)\mathcal{L}_{Bo}^{n-1}\right], i=1,...,M,
\end{equation}
where $\mathcal{L}_{Bo}^{n-1}$ is the distillation loss to calibrate the $(n-1)$-th sub-model computed by aligning it with the large transformer.

\textbf{Model distillation.} 
The updated weights are then used in the distillation process. 
Following the approach in DeiT \cite{deit}, knowledge from the large transformer is transferred to the current sub-model.
For the $n$-th iteration, the objective function of distillation is defined as
\begin{equation}
  \label{eq:boost_loss}
  \mathcal{L}_{Bo}^{n} = \frac{W_{n}}{2} \left[\mathcal{L}_{CE}(\varLambda (Y_{s_n}),y)+\mathcal{L}_{CE}(\varLambda(Y_{s_n}),y_t)\right],
\end{equation}
where $\varLambda(\cdot)$ is the softmax function and $\mathcal{L}_{CE}$ is the cross entropy loss.
The prediction logits of the large transformer and $n$-th sub-model are denoted as $Y_t$ and $Y_{s_n}$, respectively.  
$y_t$ is the hard decision of the large transformer.
This process is repeated until all sub-models are sequentially calibrated.

\textbf{Discussion.}
Traditional ensemble learning methods typically need to train all models independently from scratch and then aggregate their predictions to generate the final output.
In contrast, CoFormer decomposes a pre-trained large transformer into multiple sub-models. These sub-models are calibrated by employing a progressive distillation process to transfer knowledge from the original large transformer model.
These calibrated sub-models perform inference in parallel on different edge devices. Their intermediate results from these devices are aggregated by the central device using an efficient aggregation module to produce the final output.
Our method significantly reduces the inference latency and communication overhead compared to traditional ensemble methods, while maintaining high accuracy.

\begin{figure}
  \centering
  \includegraphics[width=0.4\textwidth]{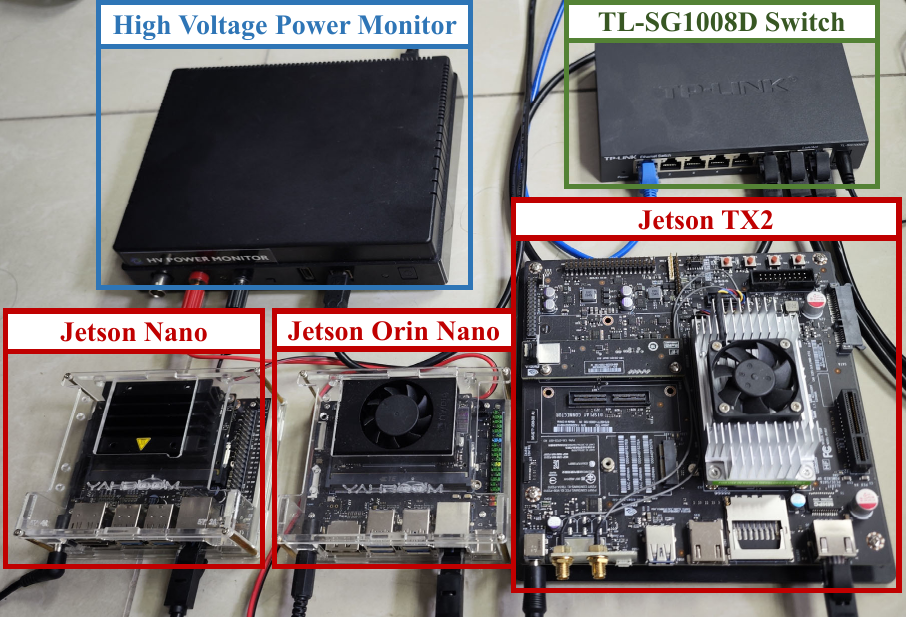}
  \caption{Our system prototype employs a Jetson Nano, a Jetson Orin Nano, a Jetson TX2 and a switch. We utilize the Monsoon High Voltage Power Monitor to measure the energy.}
  \label{fig:prototype}
\end{figure}

\begin{figure*}[ht]
  \centering
  \includegraphics[width=0.92\textwidth]{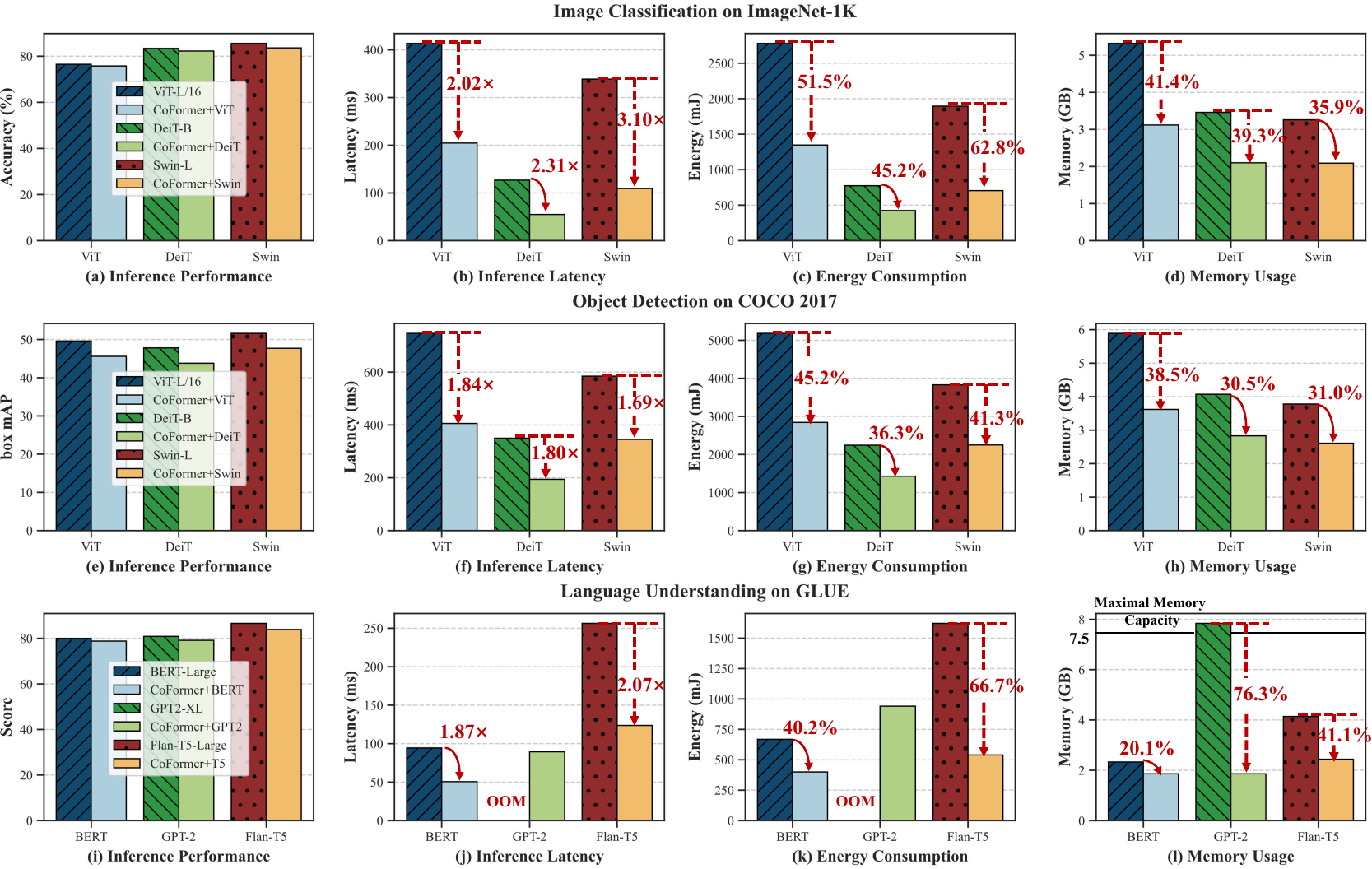}
  \vspace{-4pt}
  \caption{The evaluation results for inference performance, latency, energy consumption, and memory usage across image classification (a-d), object detection (e-h), and language understanding tasks (i-l). 
  Notably, the inference latency and energy consumption of GPT2-XL are marked as "OOM"  due to its excessive memory requirements for edge devices. 
  Our method achieves 2$\times$ faster inference speed, reduces energy consumption by over 35\%, and decreases memory usage by more than 20\%.}
  \vspace{-4pt}
  \label{fig:re_all}
\end{figure*}

\section{Evaluation}

\subsection{Experimental Setup}

\textbf{Prototype.}
We implement our system on a real-world collaborative edge computing testbed that consits of an NVIDIA Jetson Nano with 4 GB memory, an NVIDIA Jetson TX2 with 8 GB memory, an NVIDIA Jetson Orin Nano with 4 GB memory and a switch. 
These devices represent the edge devices with different hardware resources and computing capabilities. The detailed configurations of these devices are shown in the lower left part of Table \ref{tab:device_model} of supplementary material. 
Figure \ref{fig:prototype} shows the implemented hardware platform for CoFormer. These devices are connected via a gigabyte switch TP-LINK TL-SG1008D. We use the traffic control tool tc\footnote{Tc-show, \url{https://www.linux.com/tutorials/tc-show-manipulate-traffic-control-settings/}} to control bandwidth, 
which can help simulate the diverse network conditions within realistic edge environments under different network conditions to evaluate the performance.

\textbf{Transformer backbones, tasks and datasets.}
To evaluate the generality of our system on vision and language tasks, we selected three typical intelligent applications. 
\begin{itemize}
    \item \textbf{Generic-category image recognition.} The vision task aims to recognize the generic category of an image. We select three representative ViT backbones (ViT\cite{vit}, DeiT\cite{deit}, Swin\cite{swin}) and two widely-used classification datasets (CIFAR-100\footnote{CIFAR dataset, \url{https://www.cs.toronto.edu/~kriz/cifar.html}} and ImageNet-1K\footnote{ImageNet dataset, \url{https://image-net.org/}}). 
    \item \textbf{Real-time object detection.} This type of vision tasks aims to detect objects in real-time with fast processing time while maintaining a required level of accuracy. 
    We choose Mask R-CNN \cite{mask-rcnn} with ViT, DeiT and Swin backbone as the detection model in the most widely-used dataset, MS COCO 2017\footnote{MS COCO dataset, \url{https://cocodataset.org/}}, to evaluate. 
    \item \textbf{General language understanding.} The language task aims to understand natural language. We select three language transformer backbones (BERT \cite{bert}, GPT2 \cite{gpt2} and Flan-T5 \cite{flan-t5}) and a representative General Language Understanding Evaluation (GLUE) benchmark\footnote{GLUE benchmark, \url{https://gluebenchmark.com/}} to evaluate our method. 
    The GLUE benchmark consists of nine language understanding tasks and cover a diverse range of dataset sizes and degrees of difficulty.
  \end{itemize}

\textbf{Baselines.}
We compare CoFormer with state-of-the-art edge deployment methods for transformer models by implementing three types of baselines.
\textbf{\textit{(i) Comparison with large transformer models:}} 
We compare three vision transformer models (ViT-L/16, DeiT-B and Swin-L) and three language transformer models (BERT-Large, GPT2-XL and Flan-T5-Large) as the decomposed models to demonstrate the effectiveness of our methods.
\textbf{\textit{(ii) Comparison with efficient transformer models:}} 
We compare our methods with several representative lightweight transformer models (\ie single-edge solution illustrated in Figure \ref{fig:existing} (b)) to show the superiority, including
T2T-ViT\cite{t2t}, EfficientFormer \cite{efficientformer}, PoolFormer \cite{poolformer} and MobileViTv2 \cite{mobilevitv2}. 
\textbf{\textit{(iii) Comparison with collaborative inference methods:}}
\textit{DeViT} \cite{devit} is the most relevant baseline to our work, designed for vision transformer inference on multiple homogeneous edge devices. 
\textit{Galaxy} \cite{galaxy} employs tensor parallel across attention and MLP modules for collaborative inference on edge devices.
\textit{DeTransformer} \cite{detransformer} combine block parallel and tensor parallel techniques to balance inference accuracy and communication cost. 
\textit{EdgeShard} \cite{edgeshard} splits transformer into a pipeline architecture to collaborate with edge devices.

\textbf{Evaluation metrics.}
We employ \textit{Top-1 classification accuracy}, \textit{mean average precision (mAP)} and \textit{evaluation scores on GLUE benchmark} as the overall performance metric of image classification, object detection and language understanding tasks, respectively. 
The \textit{end-to-end inference latency} and \textit{energy consumption} are utilized as the efficiency metrics. 
\textbf{\textit{(i) Measurements of inference latency:}}
We measure the end-to-end execution time in the inference process with 50 runs without break using PyTorch Profiler tool\footnote{PyTorch profiler, \url{https://pytorch.org/docs/stable/profiler.html}}. 
We record the average execution time as the end-to-end inference latency.
\textbf{\textit{(ii) Measurements of energy consumption:}}
We measure the total energy consumption of all edge devices.
We use a Monsoon High Voltage Power Monitor\footnote{Monsoon Solutions Inc., High voltage power monitor, \url{https://www.msoon.com/high-voltage-power-monitor}} connected to edge devices to obtain accurate power readings over running a forward pass of each model for 50 times. 
For each device, we derive the average energy consumption by first subtracting background power consumption (\ie power readings when not running any model) 
and calculate the integral over the individual power samples as energy consumption. The average energy consumption is recorded as the final statistics of a model. 
The measurement follows \cite{DBLP:conf/imc/AlmeidaLMDLL21} for fair comparison.

\textbf{Implementation.}
CoFormer is fully implemented with $~2,000$ LOC in Python in total atop PyTorch. 
Although we use PyTorch for auto-differentiation and computation graph execution, CoFormer is extensible and can work well with other lightweight ML frameworks such as TF-Lite and MNN.
The employed transformer models are trained following DeBo algorithm and deployed on all devices in advance. 
The communication module is implemented based on gPRC\footnote{gRPC, \url{https://grpc.io/}}. 
All the compared approaches are run with timm\footnote{timm, \url{ https://github.com/rwightman/pytorch-image-models}} (vision transformers models) or transformers\footnote{transformers, \url{https://github.com/huggingface/transformers}} (language transformer models), for fair comparison.

\subsection{End-to-End Performance}   
\label{sec:performance}

\textbf{Comparison with large transformer models.}
We compare the performance of our method with large transformers on three representative tasks including image classification, object detection and language understanding. 
The results of inference performance, inference latency, average energy consumption and memory usage are shown in Figure \ref{fig:re_all}. 

\textit{(1) Image recognition:}
The original large transformer models used for the image recognition task include ViT-L/16, DeiT-B, and Swin-L.
Since ViT-L/16 and Swin-L are impractical for deployment on Jetson Nano and Jetson Orin Nano due to their high memory demands and computational costs, we report the evaluation results of all large transformer models on Jetson TX2 for a fair comparison.
The evaluation results on ImageNet-1K and CIFAR-100 are presented in Figure \ref{fig:re_all} (a-d) and Figure \ref{fig:re_cifar} of supplementary material, respectively. 
Our method demonstrates the effectiveness of decomposing large transformer models for various backbones into smaller, more efficient models, enabling collaborative inference with constrained hardware resource requirements while maintaining satisfactory accuracy. 
Specifically, our decomposed model, based on DeiT-B, accelerates inference speed by $2.35\times$, reduce energy consumption by 45.3\% and save memory usage by 35.6\%, with only 1.14\% accuracy sacrifice on ImageNet-1K.

\textit{(2) Object detection:}
We also evaluate our system on the object detection task to validate its generality and performance. 
We utilize CoFormer to gain decomposed ViT, DeiT and Swin as backbones in Mask-RCNN pipeline and evaluate over MS COCO 2017 dataset.
These backbones are first initialized with ImageNet-1K pretrained weights and then fine-tuned on MS COCO 2017. Following the training protocol in \cite{swin,pvtv2}, we use AdamW optimizer with an initial learning rate of 1e-4 and a weight decay as 0.05. 
Note that we follow \cite{vitdet} to construct hierarchical feature maps for DeiT and ViT because they only produce a single resolution of feature maps, which is unavailable for the detection task.
The results are shown in Figure \ref{fig:re_all} (e-h). Similar to the classification task, our method can achieve fast inference speed and low energy consumption, with minimal performance degradation.

\textit{(3) Language understanding:}
We employ BERT-Large, GPT2-XL and Flan-T5-Large as transformer backbones to evaluate on the language understanding task.
Evaluation results on the GLUE benchmark, depicted in Figure \ref{fig:re_all} (i-l), demonstrates that our approach enables the effective deployment of huge transformer models on resource-constrained edge deviecs. 
Notably, our method support the GPT2-XL model, which comprises 1.6 billion model parameters, to perform efficient inference on resource-constrained edge devices, achieving a reduction in memory requirements by up to 76.3\%.
Furthermore, for BERT-Large, our approach can improve execution speed by $1.87 \times$, reduce energy consumption by 40.2\%, and decrease memory usage by 20.1\%.

\begin{table*}
  \renewcommand{\arraystretch}{0.9}
  \setlength\tabcolsep{4.5pt} 
  \centering
  \setlength{\extrarowheight}{0pt}
  \addtolength{\extrarowheight}{\aboverulesep}
  \addtolength{\extrarowheight}{\belowrulesep}
  \setlength{\aboverulesep}{0pt}
  \setlength{\belowrulesep}{0pt}
  \caption{Comparison between our method and efficient transformer models with approximate FLOPs on ImageNet-1K. }
  \vspace{-4pt}
  \label{tab:eval_baseline}
  \begin{tabular}{c|ccc|c|cc|cc} 
  \toprule
  \textbf{Method}                                                   & \textbf{FLOPs} $\downarrow$ & \textbf{Memory} $\downarrow$ & \textbf{\#Params.} $\downarrow$ & \textbf{Accuracy} $\uparrow$ & \textbf{Latency} $\downarrow$ & \textbf{Speed-up} $\uparrow$ & \textbf{Energy} $\downarrow$ & \textbf{Reduction} $\uparrow$  \\ 
  \hline
  PoolFormer-M48                                                    & 23.2 G                      & 4.39 GB                      & 56.0 M                          & 82.50 \%                     & 176.87~\scriptsize{$\pm$0.38} ms           & 1.00 $\times$               & 1156.93 \scriptsize{$\pm$86.3} mJ         & 0.00 \%                       \\
  EfficientFormer L7                                                & 20.4 G                      & 4.31 GB                      & 82.1 M                          & 83.30 \%                     & 128.84~\scriptsize{$\pm$0.33} ms           & 1.37 $\times$               & 897.46 \scriptsize{$\pm$23.1} mJ          & 22.43 \%                      \\
  T2T-ViT$_t$-19                                                    & 19.6 G                      & 2.13 GB                      & 39.2 M                          & 81.90 \%                     & 119.70~\scriptsize{$\pm$0.43} ms           & 1.48~$\times$               & 764.88 \scriptsize{$\pm$46.5} mJ          & 33.89 \%                      \\
  \rowcolor[rgb]{0.953,0.953,0.957} \textbf{\textbf{CoFormer+Swin}} & 20.3 G                      & \textbf{2.09} GB             & 68.1 M                          & \textbf{83.61} \%            & \textbf{98.30} \scriptsize{$\pm$0.97} ms   & \textbf{1.80 $\times$}      & \textbf{705.12} \scriptsize{$\pm$52.1} mJ & \textbf{39.05 \%}             \\ 
  \hline
  PoolFormer-M36                                                    & 17.6 G                      & 4.31 GB                      & 56.0 M                          & 82.10 \%                     & 134.32 \scriptsize{$\pm$0.45} ms           & 1.00 $\times$               & 858.94 \scriptsize{$\pm$63.7} mJ          & 0.00 \%                       \\
  T2T-ViT-19                                                        & 17.0 G                      & 2.12~GB                      & 39.2~M                          & 81.90 \%                     & 99.06 \scriptsize{$\pm$0.57} ms            & 1.36 $\times$               & 609.22 \scriptsize{$\pm$31.2} mJ          & 29.07 \%                      \\
  MobileViTv2-200                                                   & 15.0 G                      & 3.87 GB                      & 18.5~M                          & 81.17~\%                     & 74.32 \scriptsize{$\pm$0.84} ms            & 1.81 $\times$               & 500.92 \scriptsize{$\pm$26.9} mJ          & 41.68 \%                      \\
  \rowcolor[rgb]{0.953,0.953,0.957} \textbf{CoFormer+DeiT}          & \textbf{14.4} G             & \textbf{2.10} GB             & 37.8 M                          & \textbf{82.26} \%            & \textbf{54.89} \scriptsize{$\pm$1.62} ms   & \textbf{2.45 $\times$}      & \textbf{424.57} \scriptsize{$\pm$31.8} mJ & \textbf{50.57 \%}             \\
  \bottomrule
  \end{tabular}
  \end{table*}

  \begin{figure*}
    \centering
    \includegraphics[width=0.98\textwidth]{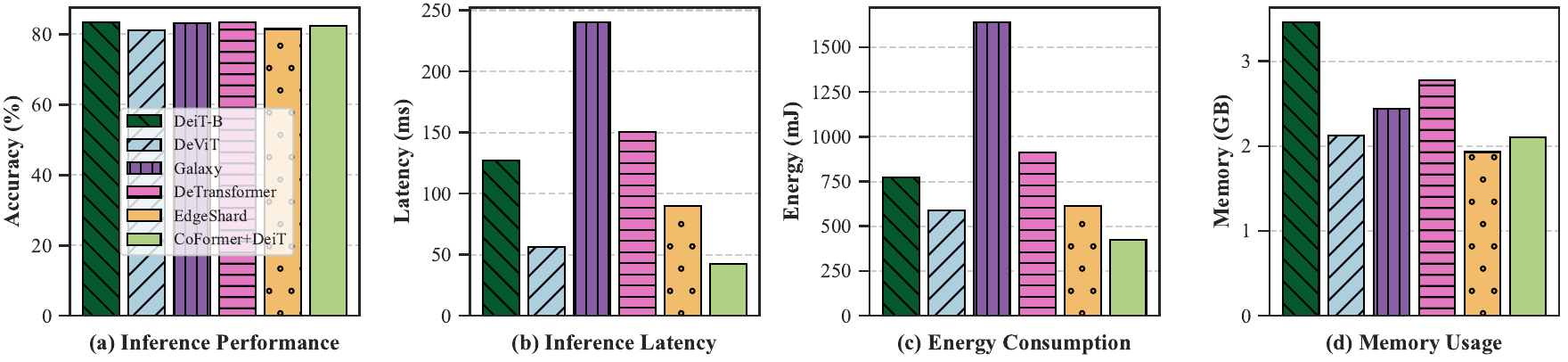}
    \caption{Performance comparison of classification accuracy (a), average inference latency (b), average energy consumption (c) and maximal memory usage (d) on ImageNet-1K using the DeiT backbone between our methods and representative collaborative inference methods.}
    \label{fig:co_re}
  \end{figure*}

\textbf{Comparison with efficient transformer models.}
Table \ref{tab:eval_baseline} presents the evaluation results between our method and efficient transformer models with similar FLOPs to ensure a fair comparison on ImageNet-1K
These models are all evaluated on the Jetson TX2 to maintain consistency. 
We can observe that our method can achieve the lowest end-to-end inference latency and energy consumption with comparable accuracy to other efficient models under varying hardware constraints.
For instance, using the DeiT backbone, our method accelerates inference speed by 26.1\%, saves energy consumption by 15.2\% and improves accuracy by 1.09\% with the lowest memory requirements and computational costs compared to state-of-the-art efficient transformer model, MobileViTv2.  

\begin{table}
  \setlength\tabcolsep{4.5pt}
  \centering
  \caption{Ablation study results of main components on CIFAR-100. The "\ding{52}" mark indicating we adopt the corresponding method. The second row represents the evaluation results of three decomposed sub-models.}
  \label{tab:ablation}
  \begin{tabular}{c|c|c|c|c|c|c|c} 
    \toprule
    \textbf{Decompose} & \textbf{Aggregate} & \multicolumn{3}{c|}{\textbf{Accuracy (\%)}} & \multicolumn{3}{c}{\textbf{Latency (ms)}}  \\ 
    \hline
    \ding{55}                 & \ding{55}                & \multicolumn{3}{c|}{91.3}                   & \multicolumn{3}{c}{123.5}                  \\ 
    \hline
    \ding{52}                 & \ding{55}                & 52.2 & 68.7 & 76.5                          & 34.2 & 25.9 & 16.4                         \\ 
    \hline
    \ding{52}                 & \ding{52}                & \multicolumn{3}{c|}{90.3}                   & \multicolumn{3}{c}{51.8}                   \\
    \bottomrule
    \end{tabular}
  \end{table}

\textbf{Comparison with collaborative inference methods.}
We implement several representative collaborative inference methods as baselines to validate the superiority of our proposed approach.
Figure \ref{fig:co_re} presents the classification accuracy, average inference latency, average energy consumption and peak memory requirements on ImageNet-1K, using the DeiT backbone. 
The results indicate CoFormer can achieve the fastest inference speed and lowest energy consumption while maintaining competitive accuracy. 
For instance, DeViT \cite{devit}, although conceptually similar, fails to consider the heterogeneity of edge devices, leading to significant synchronization delays.
While Galaxy \cite{galaxy} improves classification accuracy by 0.97\% over CoFormer but incurs an 82\% increase in inference latency due to intensive communication in its tensor-parallel design.
DeTransformer \cite{detransformer} reduces communication costs and achieves a 1.73$\times$ speedup over Galaxy, but its inference latency remains 64.6\% higher than CoFormer and comes with increased memory consumption.
EdgeShard\cite{edgeshard} exhibits lower inference speed than CoFormer due to its sequential execution strategy.
Overall, these results highlight CoFormer offers an effective trade-off between accuracy and efficiency.

\subsection{Ablation Study}

\textbf{Ablation of main components.}
We design experiments to verify the effectiveness of the proposed transformer decomposition and results aggregation. 
Table \ref{tab:ablation} presents the inference accuracy and latency utilizing different components for DeiT-B. These results show that these components of our system are all significant. 
Specifically, we can observe that transformer decomposition accelerates inference speed, leading to considerable performance degradation. 
By utilizing results aggregation, the inference efficiency and accuracy can achieve good trade-off. 

\begin{figure}
    \centering
    \includegraphics[width=0.45\textwidth]{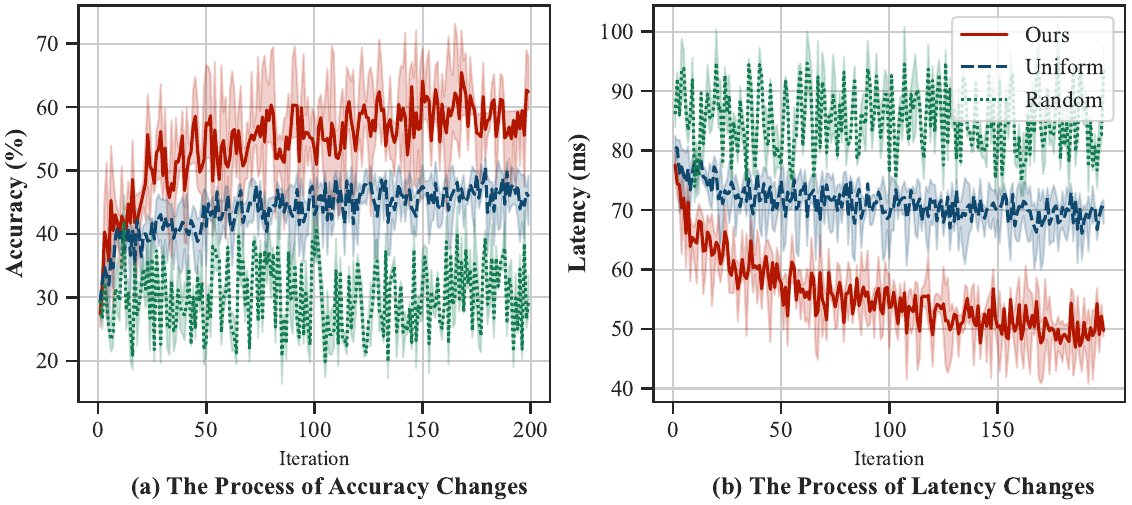}
    \vspace{-4pt}
    \caption{Accuracy (a) and latency (b) changes during the splitting process of DeiT-B on ImageNet-1K.}
    \vspace{-4pt}
    \label{fig:search}
  \end{figure}

\textbf{Transformer decomposition.}
In order to evaluate the effectiveness of the proposed transformer decomposition, we compare random decomposition and uniform decomposition with the same computational constraints. 
Uniform decomposition represents we decompose the large transformer into the multiple sub-models with the same structure. The results of accuracy and latency in the decomposition process are shown in Figure \ref{fig:search}. 
We can observe that our method performs the best with the highest accuracy and lowest inference latency. 
Although the uniform decomposition convergences faster than others, its inference latency is higher due to the considerable synchronization delay.
The random decomposition is unstable and uncontrollable, which cannot be utilized to decompose. 

\textbf{Results aggregation.}
We compare our methods with other representative aggregation methods, and they all utilize the same decomposed small transformers. The accuracy and inference latency on ImageNet-1K are shown in Table \ref{tab:agg}. 
Traditional ensemble methods, such as average and majority voting \cite{DBLP:journals/air/YangLC23}, provide fast aggregation with limited performance improvement.
In contrast, attention \cite{DBLP:conf/nips/NagraniYAJSS21} and SENet \cite{DBLP:journals/pami/HuSASW20} methods can focus on target features during aggregation.
Our method achieves the lowest inference latency with only 1.14\% accuracy sacrifice, demonstrating its ability to effectively balance latency and accuracy.

\begin{table}
    \renewcommand{\arraystretch}{0.7}
    \centering
    \setlength{\extrarowheight}{0pt}
    \addtolength{\extrarowheight}{\aboverulesep}
    \addtolength{\extrarowheight}{\belowrulesep}
    \setlength{\aboverulesep}{0pt}
    \setlength{\belowrulesep}{0pt}
    \caption{Comparison of different aggregating methods. CoFormer achieves good balance between accuracy and latency.}
    \label{tab:agg}
    \resizebox{0.82\linewidth}{!}{
      \begin{tabular}{c|cc} 
      \toprule
      \textbf{Aggregating Method}                            & \textbf{Accuracy} & \textbf{Latency}            \\ 
      \hline
      DeiT-B                                          & 83.40 \%          & 126.91 \scriptsize{$\pm$0.46} ms          \\ 
      \hline
      Average \cite{DBLP:journals/air/YangLC23}                                        & 81.02 \%          & 54.84 \scriptsize{$\pm$0.53} ms           \\
      Majority Voting \cite{DBLP:journals/air/YangLC23}                                & 80.71 \%          & 54.42 \scriptsize{$\pm$0.24} ms           \\
      Attention  \cite{DBLP:conf/nips/NagraniYAJSS21}                                     & 82.53 \%          & 61.44 \scriptsize{$\pm$0.49} ms           \\
      SENet \cite{DBLP:journals/pami/HuSASW20}                                          & 82.67 \%          & 58.74 \scriptsize{$\pm$0.38} ms           \\
      \rowcolor[rgb]{0.953,0.953,0.957} \textbf{CoFormer} & 82.26 \%          & \textbf{54.89} \scriptsize{$\pm$0.62} ms  \\
      \bottomrule
      \end{tabular}
    }
    \end{table}

\begin{figure}
  \centering
  \includegraphics[width=0.49\textwidth]{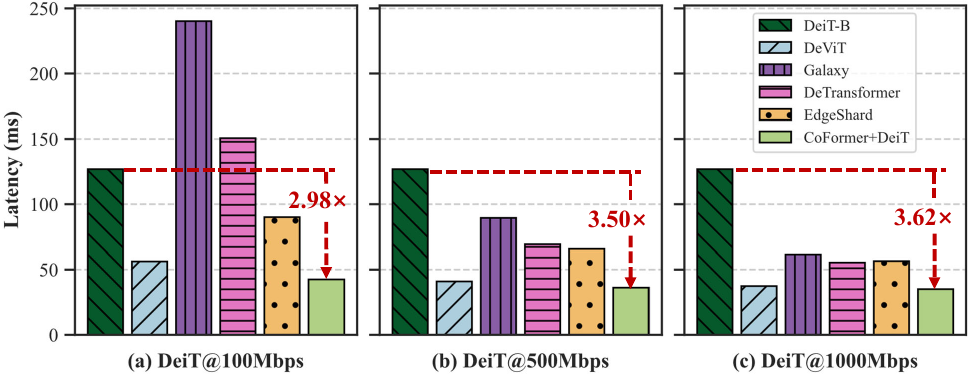}
  \caption{
    End-to-end inference latency using the DeiT backbone on CIFAR-100 across different network bandwidths, \ie 100 Mbps (a), 500 Mbps (b), and 1 Gbps (c).
    Our method demonstrates substantial acceleration across bandwidth conditions, with up to 3.62$\times$ speedup at 1 Gbps.
  }
  \label{fig:co_band}
\end{figure}

\textbf{Impact of network bandwidths. }
We evaluate the end-to-end inference latency using the DeiT backbone on CIFAR-100 across different network bandwidths, \ie 100 Mbps, 500 Mbps, and 1 Gbps, as illustrated in Figure \ref{fig:co_band}.
The inference latency of DeiT-B is measured on the Jetson TX2 platform.
The results show that our approach consistently reduces inference latency across all tested bandwidths. Notably, higher bandwidths alleviate communication overhead, further amplifying the performance gains.
For instance, CoFormer achieves $2.98\times$ speedup at 100 Mbps, and the improvement increases to $3.62\times$ at 1 Gbps compared to DeiT-B.
Our method outperforms other collaborative inference baselines and achieves the lowest inference latency across different bandwidths.
Specifically, CoFormer accelerates inference speed by $5.65\times$ speedup at 100 Mbps and $1.76\times$ at 1 Gbps compared to Galaxy.
These findings highlight the robustness and effectiveness of our method across diverse network environments.

\begin{figure}[t]
  \centering
  \includegraphics[width=0.48\textwidth]{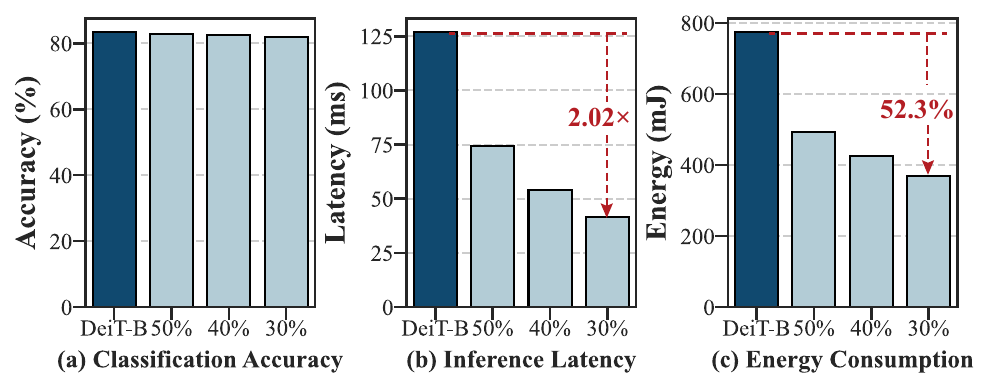}
  \caption{Results for DeiT backbone on ImageNet-1K under different constraints. The percentage represents the ratio of computational constraints.}
  \label{fig:constraints}
\end{figure}

\textbf{Impact of resource constraints.}
Figure \ref{fig:constraints} illustrates the performance of our method for DeiT-B under several resource constraints. 
We impose computational limits of 30\%, 40\%, and 50\% for edge deployments. 
At a computational cost of 30\%, our method improves inference speed by 3.05$\times$ and reduces energy consumption by 52.3\%, with only a slight accuracy sacrifice of 1.56\%. 
Given that accuracy losses up to 2\% are acceptable for substantial latency improvements in most real-world applications, it demonstrates the flexibility and efficiency of our method under varied resource constraints.

\textbf{Impact of device quantity.}
We examine the impact of altering the quantity of devices on inference performance, as depicted in Table \ref{tab:device_combine}. 
The reported energy consumption is the total energy consumed across all devices. 
To ensure a fair comparison, the total FLOPs across devices are kept consistent.
As the number of devices increases, a slight reduction in accuracy is observed, not surpassing a 2\% decrease, demonstrating the robustness of our method to changes in device quantity.
CoFormer achieves lower inference latency and total energy consumption using multiple edge devices compared to single-device inference for large transformer models. 
These benefits stem from parallel execution, which shortens overall inference time and reduces energy consumption, even with more devices involved.
However, the marginal gains in latency and energy diminish as the device count grows, due to increasing communication and idle overheads.

\begin{table}[t]
  \centering
  \caption{Results of different device number on CIFAR-100. "Num.=1" represents the inference of DeiT-B in Jetson TX2.}
  \label{tab:device_combine}
    \begin{tabular}{c|ccc} 
    \toprule
    \textbf{Num.} & \textbf{Accuracy} & \textbf{Latency}    & \textbf{\textbf{Energy}}  \\ 
    \hline
    1             & 91.30 \%            & 123.48 \scriptsize{$\pm$0.43} ms          & 760.64 \scriptsize{$\pm$57.3} mJ        \\ 
    \hline
    2             & 90.36 \%          & 85.64 \scriptsize{$\pm$0.52} ms           & 618.94 \scriptsize{$\pm$38.2} mJ       \\ 
    \hline
    3             & 90.27 \%          & 51.84 \scriptsize{$\pm$0.81} ms           & 409.55 \scriptsize{$\pm$42.7} mJ       \\ 
    \hline
    4             & 90.03 \%          & 45.48 \scriptsize{$\pm$0.65} ms           & 367.46 \scriptsize{$\pm$24.1} mJ       \\
    \bottomrule
    \end{tabular}
  \end{table}

\section{Related Work}
\textbf{Transformer deployment.}
Deploying transformers on resource-constrained edge devices is challenging due to substantial computational requirements. Recently several works designed lightweight architectures for intelligent applications on such devices.
Maaz \textit{et al.} \cite{mobilevit} introduced a convolutional structure into transformers to leverage both convolutional and attention mechanisms, developing an attention mechanism with linear complexity \cite{mobilevitv2}. 
On the other hand, various methods have employed model compression techniques, including network pruning \cite{DBLP:conf/cvpr/YuX23, DBLP:conf/cvpr/YuCGF23}, knowledge distillation \cite{DBLP:conf/nips/HaoG0TH0X23, deit}, low-bit quantization \cite{DBLP:conf/cvpr/YuCGF23} and NAS \cite{autoformerv2, ofa} to gain compact transformer models.  
For instance, Hao \textit{et al.} \cite{DBLP:conf/nips/HaoG0TH0X23} projected intermediate features into an aligned latent space to help heterogeneous student models imitate teacher models.

\textbf{Collaborative inference.}
This technique involves dividing a DNN into multiple segments for deployment across different devices to facilitate inference. Recent research in collaborative inference falls into two categories: 
\textit{(i) Layer-wise splitting. }
A DNN is partitioned by layer into multiple groups and distributed across edge nodes, sometimes involving direct transmission of input data to the cloud for inference.
Most work focuses on finding an optimal partitioning point and scheduling strategy for balancing performance and efficiency \cite{DBLP:journals/tpds/YangZZSGL22, DBLP:journals/tmc/HaoXLHAM23, DBLP:conf/mobicom/HuangG22, DBLP:journals/ijautcomp/RenQ0JSW023}. 
Yang \textit{et al.} \cite{DBLP:journals/tpds/YangZZSGL22} proposed a deep reinforcement learning algorithm for optimizing task offloading strategies.
\textit{(ii) Across-layer splitting.} 
It involves dividing the DNN weights into multiple unconnected groups for parallel execution on various edge devices \cite{devit, DBLP:journals/tmc/TuliCJ23, DBLP:journals/tpds/ZhouLWMW23,galaxy,detransformer,wei2024communication}.
These methods require an additional training procedure where publicly available pre-trained models cannot be used after splitting, compared to layer-wise splitting. 
Tuli \textit{et al.} \cite{DBLP:journals/tmc/TuliCJ23} utilized Multi-Armed-Bandits for determining the optimal division and distribution of DNNs across specific devices.

Existing strategies for deploying transformer models primarily focus on compressing models or designing lightweight architectures tailored for individual edge devices. In this work, we explore an alternative direction by leveraging the inherent divisibility and integrability of transformer architectures to enable collaborative inference across multiple edge devices. 
Specifically, we propose to decompose large transformer models into segments that are distributed for parallel execution, thereby reducing the per-device computational burden.
Among existing collaborative inference techniques, the most closely related approaches are Galaxy \cite{galaxy}, DeTransformer \cite{detransformer, wei2024communication} and DeViT \cite{devit}.
Galaxy adopts a tensor-parallel strategy by partitioning the weights of attention and MLP modules across devices. DeTransformer employs block-parallel execution to alleviate communication overhead. However, both methods require multiple rounds of inter-device communication during inference, which introduces significant latency. 
In contrast, our approach requires only a single round of communication for final result aggregation, substantially reducing communication costs.
DeViT decomposes vision transformer for distributed inference on multiple homogeneous edge devices.
However, to the best of our knowledge, CoFormer is the first work to automatically decompose off-the-shelf large transformer models for collaborative inference across heterogeneous edge devices to achieve a good balance between efficiency and accuracy.

\section{Conclusion}
In this paper, we propose a collaborative inference system for general transformers to facilitate efficient deployment across heterogeneous edge devices.
An off-the-shelf large transformer can be automatically decomposed into multiple smaller models tailored to the specific hardware constraints.
To balance inference performance and efficiency, we formulate an optimization problem to minimize both inference latency and accuracy degradation.
We also develop the DeBo algorithm to address this optimization problem, leveraging Bayesian optimization to derive an optimal decomposition policy and progressively calibrate sub-models.
Extensive experiments demonstrates our system can accelerate inference speed by 1.7$\times$ to 3.1$\times$ and reduces energy consumption by 36.3\% to 63.8\% with satisfactory performance.

\newpage

\bibliographystyle{IEEEtran}
\bibliography{ref.bib}

\begin{thebibliography}{10}
\providecommand{\url}[1]{#1}
\csname url@samestyle\endcsname
\providecommand{\newblock}{\relax}
\providecommand{\bibinfo}[2]{#2}
\providecommand{\BIBentrySTDinterwordspacing}{\spaceskip=0pt\relax}
\providecommand{\BIBentryALTinterwordstretchfactor}{4}
\providecommand{\BIBentryALTinterwordspacing}{\spaceskip=\fontdimen2\font plus
\BIBentryALTinterwordstretchfactor\fontdimen3\font minus \fontdimen4\font\relax}
\providecommand{\BIBforeignlanguage}[2]{{%
\expandafter\ifx\csname l@#1\endcsname\relax
\typeout{** WARNING: IEEEtran.bst: No hyphenation pattern has been}%
\typeout{** loaded for the language `#1'. Using the pattern for}%
\typeout{** the default language instead.}%
\else
\language=\csname l@#1\endcsname
\fi
#2}}
\providecommand{\BIBdecl}{\relax}
\BIBdecl

\bibitem{DBLP:conf/iclr/ChenMCWP025}
Y.~Chen, M.~Mihajlovic, X.~Chen, Y.~Wang, S.~Prokudin, and S.~Tang, ``Splatformer: Point transformer for robust 3d gaussian splatting,'' in \emph{The Thirteenth International Conference on Learning Representations, {ICLR} 2025, Singapore, April 24-28, 2025}.\hskip 1em plus 0.5em minus 0.4em\relax OpenReview.net, 2025.

\bibitem{DBLP:journals/pami/ChittaPJYRG23}
K.~Chitta, A.~Prakash, B.~Jaeger, Z.~Yu, K.~Renz, and A.~Geiger, ``Transfuser: Imitation with transformer-based sensor fusion for autonomous driving,'' \emph{{IEEE} Trans. Pattern Anal. Mach. Intell.}, vol.~45, no.~11, pp. 12\,878--12\,895, 2023.

\bibitem{DBLP:conf/cvpr/MaSZ0XYY22}
F.~Ma, M.~Z. Shou, L.~Zhu, H.~Fan, Y.~Xu, Y.~Yang, and Z.~Yan, ``Unified transformer tracker for object tracking,'' in \emph{{IEEE/CVF} Conference on Computer Vision and Pattern Recognition, {CVPR} 2022, New Orleans, LA, USA, June 18-24, 2022}.\hskip 1em plus 0.5em minus 0.4em\relax {IEEE}, 2022, pp. 8771--8780.

\bibitem{swin}
Z.~Liu, Y.~Lin, Y.~Cao, H.~Hu, Y.~Wei, Z.~Zhang, S.~Lin, and B.~Guo, ``Swin transformer: Hierarchical vision transformer using shifted windows,'' in \emph{2021 {IEEE/CVF} International Conference on Computer Vision, {ICCV} 2021, Montreal, QC, Canada, October 10-17, 2021}.\hskip 1em plus 0.5em minus 0.4em\relax {IEEE}, 2021, pp. 9992--10\,002.

\bibitem{flan-t5}
H.~W. Chung, L.~Hou, S.~Longpre, B.~Zoph, Y.~Tay, W.~Fedus, Y.~Li, X.~Wang, M.~Dehghani, S.~Brahma \emph{et~al.}, ``Scaling instruction-finetuned language models,'' \emph{Journal of Machine Learning Research}, vol.~25, no.~70, pp. 1--53, 2024.

\bibitem{gpt2}
A.~Radford, J.~Wu, R.~Child, D.~Luan, D.~Amodei, I.~Sutskever \emph{et~al.}, ``Language models are unsupervised multitask learners,'' \emph{OpenAI blog}, vol.~1, no.~8, p.~9, 2019.

\bibitem{DBLP:journals/tkde/YaoZYWMZCJJSWZTKWWZY23}
J.~Yao, S.~Zhang, Y.~Yao, F.~Wang, J.~Ma, J.~Zhang, Y.~Chu, L.~Ji, K.~Jia, T.~Shen, A.~Wu, F.~Zhang, Z.~Tan, K.~Kuang, C.~Wu, F.~Wu, J.~Zhou, and H.~Yang, ``Edge-cloud polarization and collaboration: {A} comprehensive survey for {AI},'' \emph{{IEEE} Trans. Knowl. Data Eng.}, vol.~35, no.~7, pp. 6866--6886, 2023.

\bibitem{DBLP:journals/tpds/YangZZSGL22}
S.~Yang, Z.~Zhang, C.~Zhao, X.~Song, S.~Guo, and H.~Li, ``{CNNPC:} end-edge-cloud collaborative {CNN} inference with joint model partition and compression,'' \emph{{IEEE} Trans. Parallel Distributed Syst.}, vol.~33, no.~10, pp. 4039--4056, 2022.

\bibitem{DBLP:journals/tmc/HaoXLHAM23}
Z.~Hao, G.~Xu, Y.~Luo, H.~Hu, J.~An, and S.~Mao, ``Multi-agent collaborative inference via {DNN} decoupling: Intermediate feature compression and edge learning,'' \emph{{IEEE} Trans. Mob. Comput.}, vol.~22, no.~10, pp. 6041--6055, 2023.

\bibitem{DBLP:conf/mobicom/HuangG22}
K.~Huang and W.~Gao, ``Real-time neural network inference on extremely weak devices: agile offloading with explainable {AI},'' in \emph{{ACM} MobiCom '22: The 28th Annual International Conference on Mobile Computing and Networking, Sydney, NSW, Australia, October 17 - 21, 2022}.\hskip 1em plus 0.5em minus 0.4em\relax {ACM}, 2022, pp. 200--213.

\bibitem{DBLP:journals/ijautcomp/RenQ0JSW023}
W.~Ren, Y.~Qu, C.~Dong, Y.~Jing, H.~Sun, Q.~Wu, and S.~Guo, ``A survey on collaborative {DNN} inference for edge intelligence,'' \emph{Int. J. Autom. Comput.}, vol.~20, no.~3, pp. 370--395, 2023.

\bibitem{3gpp.22.874}
3GPP, ``3rd generation partnership project; {Technical} specification group services and system aspects; {Study} on traffic characteristics and performance requirements for {AI/ML} model transfer in {5GS}; ({Release} 18),'' {3rd Generation Partnership Project (3GPP)}, Technical Specification (TS) 22.874, Dec. 2021, version 18.2.0.

\bibitem{deepthing}
Z.~Zhao, K.~M. Barijough, and A.~Gerstlauer, ``Deepthings: Distributed adaptive deep learning inference on resource-constrained iot edge clusters,'' \emph{{IEEE} Trans. Comput. Aided Des. Integr. Circuits Syst.}, vol.~37, no.~11, pp. 2348--2359, 2018.

\bibitem{coedge}
L.~Zeng, X.~Chen, Z.~Zhou, L.~Yang, and J.~Zhang, ``Coedge: Cooperative {DNN} inference with adaptive workload partitioning over heterogeneous edge devices,'' \emph{{IEEE/ACM} Trans. Netw.}, vol.~29, no.~2, pp. 595--608, 2021.

\bibitem{galaxy}
S.~Ye, J.~Du, L.~Zeng, W.~Ou, X.~Chu, Y.~Lu, and X.~Chen, ``Galaxy: {A} resource-efficient collaborative edge {AI} system for in-situ transformer inference,'' in \emph{{IEEE} {INFOCOM} 2024 - {IEEE} Conference on Computer Communications, Vancouver, BC, Canada, May 20-23, 2024}.\hskip 1em plus 0.5em minus 0.4em\relax {IEEE}, 2024, pp. 1001--1010.

\bibitem{mobilevit}
S.~Mehta and M.~Rastegari, ``Mobilevit: Light-weight, general-purpose, and mobile-friendly vision transformer,'' in \emph{The Tenth International Conference on Learning Representations, {ICLR} 2022, Virtual Event, April 25-29, 2022}.\hskip 1em plus 0.5em minus 0.4em\relax OpenReview.net, 2022.

\bibitem{mobilevitv2}
{S. Mehta and M. Rastegari}, ``Separable self-attention for mobile vision transformers,'' \emph{Trans. Mach. Learn. Res.}, vol. 2023, 2023.

\bibitem{deit}
H.~Touvron, M.~Cord, M.~Douze, F.~Massa, A.~Sablayrolles, and H.~J{\'{e}}gou, ``Training data-efficient image transformers {\&} distillation through attention,'' in \emph{Proceedings of the 38th International Conference on Machine Learning, {ICML} 2021, 18-24 July 2021, Virtual Event}, ser. Proceedings of Machine Learning Research, vol. 139.\hskip 1em plus 0.5em minus 0.4em\relax {PMLR}, 2021, pp. 10\,347--10\,357.

\bibitem{DBLP:conf/cvpr/YuX23}
L.~Yu and W.~Xiang, ``X-pruner: explainable pruning for vision transformers,'' in \emph{{IEEE/CVF} Conference on Computer Vision and Pattern Recognition, {CVPR} 2023, Vancouver, BC, Canada, June 17-24, 2023}.\hskip 1em plus 0.5em minus 0.4em\relax {IEEE}, 2023, pp. 24\,355--24\,363.

\bibitem{DBLP:conf/cvpr/YuCGF23}
C.~Yu, T.~Chen, Z.~Gan, and J.~Fan, ``Boost vision transformer with gpu-friendly sparsity and quantization,'' in \emph{{IEEE/CVF} Conference on Computer Vision and Pattern Recognition, {CVPR} 2023, Vancouver, BC, Canada, June 17-24, 2023}.\hskip 1em plus 0.5em minus 0.4em\relax {IEEE}, 2023, pp. 22\,658--22\,668.

\bibitem{autoformerv2}
M.~Chen, K.~Wu, B.~Ni, H.~Peng, B.~Liu, J.~Fu, H.~Chao, and H.~Ling, ``Searching the search space of vision transformer,'' in \emph{Annual Conference on Neural Information Processing Systems 2021, NeurIPS 2021, December 6-14, 2021, virtual}, 2021, pp. 8714--8726.

\bibitem{ofa}
H.~Cai, C.~Gan, T.~Wang, Z.~Zhang, and S.~Han, ``Once-for-all: Train one network and specialize it for efficient deployment,'' in \emph{8th International Conference on Learning Representations, {ICLR} 2020, Addis Ababa, Ethiopia, April 26-30, 2020}.\hskip 1em plus 0.5em minus 0.4em\relax OpenReview.net, 2020.

\bibitem{vit}
A.~Dosovitskiy, L.~Beyer, A.~Kolesnikov, D.~Weissenborn, X.~Zhai, T.~Unterthiner, M.~Dehghani, M.~Minderer, G.~Heigold, S.~Gelly, J.~Uszkoreit, and N.~Houlsby, ``An image is worth 16x16 words: Transformers for image recognition at scale,'' in \emph{9th International Conference on Learning Representations, {ICLR} 2021, Virtual Event, Austria, May 3-7, 2021}.\hskip 1em plus 0.5em minus 0.4em\relax OpenReview.net, 2021.

\bibitem{bert}
J.~Devlin, M.~Chang, K.~Lee, and K.~Toutanova, ``{BERT:} pre-training of deep bidirectional transformers for language understanding,'' in \emph{Proceedings of the 2019 Conference of the North American Chapter of the Association for Computational Linguistics: Human Language Technologies, {NAACL-HLT} 2019, Minneapolis, MN, USA, June 2-7, 2019, Volume 1 (Long and Short Papers)}, J.~Burstein, C.~Doran, and T.~Solorio, Eds.\hskip 1em plus 0.5em minus 0.4em\relax Association for Computational Linguistics, 2019, pp. 4171--4186.

\bibitem{DBLP:journals/tpds/ZhouLWMW23}
H.~Zhou, M.~Li, N.~Wang, G.~Min, and J.~Wu, ``Accelerating deep learning inference via model parallelism and partial computation offloading,'' \emph{{IEEE} Trans. Parallel Distributed Syst.}, vol.~34, no.~2, pp. 475--488, 2023.

\bibitem{efficientformer}
Y.~Li, G.~Yuan, Y.~Wen, J.~Hu, G.~Evangelidis, S.~Tulyakov, Y.~Wang, and J.~Ren, ``Efficientformer: Vision transformers at mobilenet speed,'' in \emph{Annual Conference on Neural Information Processing Systems 2022, NeurIPS 2022, New Orleans, LA, USA, November 28 - December 9, 2022}, 2022.

\bibitem{DBLP:conf/acl/VoitaTMST19}
E.~Voita, D.~Talbot, F.~Moiseev, R.~Sennrich, and I.~Titov, ``Analyzing multi-head self-attention: Specialized heads do the heavy lifting, the rest can be pruned,'' in \emph{Proceedings of the 57th Conference of the Association for Computational Linguistics, {ACL} 2019, Florence, Italy, July 28- August 2, 2019, Volume 1: Long Papers}.\hskip 1em plus 0.5em minus 0.4em\relax Association for Computational Linguistics, 2019, pp. 5797--5808.

\bibitem{DBLP:conf/blackboxnlp/ClarkKLM19}
K.~Clark, U.~Khandelwal, O.~Levy, and C.~D. Manning, ``What does {BERT} look at? an analysis of bert's attention,'' in \emph{Proceedings of the 2019 {ACL} Workshop BlackboxNLP: Analyzing and Interpreting Neural Networks for NLP, Florence, Italy, August 1, 2019}.\hskip 1em plus 0.5em minus 0.4em\relax Association for Computational Linguistics, 2019, pp. 276--286.

\bibitem{DBLP:conf/cvpr/KhakzarBK00N21}
A.~Khakzar, S.~Baselizadeh, S.~Khanduja, C.~Rupprecht, S.~T. Kim, and N.~Navab, ``Neural response interpretation through the lens of critical pathways,'' in \emph{{IEEE} Conference on Computer Vision and Pattern Recognition, {CVPR} 2021, virtual, June 19-25, 2021}.\hskip 1em plus 0.5em minus 0.4em\relax Computer Vision Foundation / {IEEE}, 2021, pp. 13\,528--13\,538.

\bibitem{DBLP:journals/air/YangLC23}
Y.~Yang, H.~Lv, and N.~Chen, ``A survey on ensemble learning under the era of deep learning,'' \emph{Artif. Intell. Rev.}, vol.~56, no.~6, pp. 5545--5589, 2023.

\bibitem{DBLP:journals/cj/ChadhaK21}
A.~Chadha and B.~Kaushik, ``A survey on prediction of suicidal ideation using machine and ensemble learning,'' \emph{Comput. J.}, vol.~64, no.~11, pp. 1617--1632, 2021.

\bibitem{mask-rcnn}
K.~He, G.~Gkioxari, P.~Doll{\'{a}}r, and R.~B. Girshick, ``Mask {R-CNN},'' in \emph{{IEEE} International Conference on Computer Vision, {ICCV} 2017, Venice, Italy, October 22-29, 2017}.\hskip 1em plus 0.5em minus 0.4em\relax {IEEE} Computer Society, 2017, pp. 2980--2988.

\bibitem{t2t}
L.~Yuan, Y.~Chen, T.~Wang, W.~Yu, Y.~Shi, Z.~Jiang, F.~E.~H. Tay, J.~Feng, and S.~Yan, ``Tokens-to-token vit: Training vision transformers from scratch on imagenet,'' in \emph{2021 {IEEE/CVF} International Conference on Computer Vision, {ICCV} 2021, Montreal, QC, Canada, October 10-17, 2021}.\hskip 1em plus 0.5em minus 0.4em\relax {IEEE}, 2021, pp. 538--547.

\bibitem{poolformer}
W.~Yu, M.~Luo, P.~Zhou, C.~Si, Y.~Zhou, X.~Wang, J.~Feng, and S.~Yan, ``Metaformer is actually what you need for vision,'' in \emph{{IEEE/CVF} Conference on Computer Vision and Pattern Recognition, {CVPR} 2022, New Orleans, LA, USA, June 18-24, 2022}.\hskip 1em plus 0.5em minus 0.4em\relax {IEEE}, 2022, pp. 10\,809--10\,819.

\bibitem{devit}
G.~Xu, Z.~Hao, Y.~Luo, H.~Hu, J.~An, and S.~Mao, ``Devit: Decomposing vision transformers for collaborative inference in edge devices,'' \emph{{IEEE} Trans. Mob. Comput.}, vol.~23, no.~5, pp. 5917--5932, 2024.

\bibitem{detransformer}
J.~Du, Y.~Wei, S.~Ye, J.~Jiang, X.~Chen, D.~Huang, and Y.~Lu, ``Co-designing transformer architectures for distributed inference with low communication,'' \emph{IEEE Transactions on Parallel and Distributed Systems}, 2025.

\bibitem{edgeshard}
M.~Zhang, X.~Shen, J.~Cao, Z.~Cui, and S.~Jiang, ``Edgeshard: Efficient llm inference via collaborative edge computing,'' \emph{IEEE Internet of Things Journal}, 2024.

\bibitem{DBLP:conf/imc/AlmeidaLMDLL21}
M.~Almeida, S.~Laskaridis, A.~Mehrotra, L.~Dudziak, I.~Leontiadis, and N.~D. Lane, ``Smart at what cost?: characterising mobile deep neural networks in the wild,'' in \emph{{IMC} '21: {ACM} Internet Measurement Conference, Virtual Event, USA, November 2-4, 2021}, D.~Levin, A.~Mislove, J.~Amann, and M.~Luckie, Eds.\hskip 1em plus 0.5em minus 0.4em\relax {ACM}, 2021, pp. 658--672.

\bibitem{pvtv2}
W.~Wang, E.~Xie, X.~Li, D.-P. Fan, K.~Song, D.~Liang, T.~Lu, P.~Luo, and L.~Shao, ``Pvtv2: Improved baselines with pyramid vision transformer,'' \emph{Computational Visual Media}, vol.~8, no.~3, pp. 1--10, 2022.

\bibitem{vitdet}
Y.~Li, H.~Mao, R.~B. Girshick, and K.~He, ``Exploring plain vision transformer backbones for object detection,'' in \emph{Computer Vision - {ECCV} 2022 - 17th European Conference, Tel Aviv, Israel, October 23-27, 2022, Proceedings, Part {IX}}, ser. Lecture Notes in Computer Science, S.~Avidan, G.~J. Brostow, M.~Ciss{\'{e}}, G.~M. Farinella, and T.~Hassner, Eds., vol. 13669.\hskip 1em plus 0.5em minus 0.4em\relax Springer, 2022, pp. 280--296.

\bibitem{DBLP:conf/nips/NagraniYAJSS21}
A.~Nagrani, S.~Yang, A.~Arnab, A.~Jansen, C.~Schmid, and C.~Sun, ``Attention bottlenecks for multimodal fusion,'' in \emph{Annual Conference on Neural Information Processing Systems 2021, NeurIPS 2021, December 6-14, 2021, virtual}, 2021, pp. 14\,200--14\,213.

\bibitem{DBLP:journals/pami/HuSASW20}
J.~Hu, L.~Shen, S.~Albanie, G.~Sun, and E.~Wu, ``Squeeze-and-excitation networks,'' \emph{{IEEE} Trans. Pattern Anal. Mach. Intell.}, vol.~42, no.~8, pp. 2011--2023, 2020.

\bibitem{DBLP:conf/nips/HaoG0TH0X23}
Z.~Hao, J.~Guo, K.~Han, Y.~Tang, H.~Hu, Y.~Wang, and C.~Xu, ``One-for-all: Bridge the gap between heterogeneous architectures in knowledge distillation,'' in \emph{Annual Conference on Neural Information Processing Systems 2023, NeurIPS 2023, New Orleans, LA, USA, December 10 - 16, 2023}, 2023.

\bibitem{DBLP:journals/tmc/TuliCJ23}
S.~Tuli, G.~Casale, and N.~R. Jennings, ``Splitplace: {AI} augmented splitting and placement of large-scale neural networks in mobile edge environments,'' \emph{{IEEE} Trans. Mob. Comput.}, vol.~22, no.~9, pp. 5539--5554, 2023.

\bibitem{wei2024communication}
Y.~Wei, S.~Ye, J.~Jiang, X.~Chen, D.~Huang, J.~Du, and Y.~Lu, ``Communication-efficient model parallelism for distributed in-situ transformer inference,'' in \emph{2024 Design, Automation \& Test in Europe Conference \& Exhibition (DATE)}.\hskip 1em plus 0.5em minus 0.4em\relax IEEE, 2024, pp. 1--6.

\bibitem{pi}
\BIBentryALTinterwordspacing
R.~Pi, ``Raspberry pi 4 model b.'' [Online]. Available: \url{https://www.raspberrypi.com/products/raspberry-pi-4-model-b/specifications/}
\BIBentrySTDinterwordspacing

\bibitem{jetson}
\BIBentryALTinterwordspacing
NVIDIA, ``Nvidia jetson: The ai platform for edge computing.'' [Online]. Available: \url{https://www.nvidia.com/en-us/autonomousmachines/embedded-systems/}
\BIBentrySTDinterwordspacing

\bibitem{DBLP:journals/pieee/ShahriariSWAF16}
B.~Shahriari, K.~Swersky, Z.~Wang, R.~P. Adams, and N.~de~Freitas, ``Taking the human out of the loop: {A} review of bayesian optimization,'' \emph{Proc. {IEEE}}, vol. 104, no.~1, pp. 148--175, 2016.

\bibitem{williams2006gaussian}
C.~K. Williams and C.~E. Rasmussen, \emph{Gaussian processes for machine learning}.\hskip 1em plus 0.5em minus 0.4em\relax MIT press Cambridge, MA, 2006, vol.~2, no.~3.

\bibitem{DBLP:journals/jmlr/Genton01}
M.~G. Genton, ``Classes of kernels for machine learning: {A} statistics perspective,'' \emph{J. Mach. Learn. Res.}, vol.~2, pp. 299--312, 2001.

\bibitem{jones1998efficient}
D.~R. Jones, M.~Schonlau, and W.~J. Welch, ``Efficient global optimization of expensive black-box functions,'' \emph{Journal of Global optimization}, vol.~13, pp. 455--492, 1998.

\bibitem{tc}
\BIBentryALTinterwordspacing
T.~L. Foundation, ``Tc-show / manipulate traffic control settings,'' 2024. [Online]. Available: \url{https: //www.linux.com/tutorials/tc-show-manipulate-traffic-control-settings/}
\BIBentrySTDinterwordspacing

\bibitem{cifar}
A.~Krizhevsky, G.~Hinton \emph{et~al.}, ``Learning multiple layers of features from tiny images,'' 2009.

\bibitem{imagenet_1k}
J.~Deng, W.~Dong, R.~Socher, L.~Li, K.~Li, and L.~Fei{-}Fei, ``Imagenet: {A} large-scale hierarchical image database,'' in \emph{2009 {IEEE} Computer Society Conference on Computer Vision and Pattern Recognition {(CVPR} 2009), 20-25 June 2009, Miami, Florida, {USA}}.\hskip 1em plus 0.5em minus 0.4em\relax {IEEE} Computer Society, 2009, pp. 248--255.

\bibitem{coco}
T.~Lin, M.~Maire, S.~J. Belongie, J.~Hays, P.~Perona, D.~Ramanan, P.~Doll{\'{a}}r, and C.~L. Zitnick, ``Microsoft {COCO:} common objects in context,'' in \emph{Computer Vision - {ECCV} 2014 - 13th European Conference, Zurich, Switzerland, September 6-12, 2014, Proceedings, Part {V}}, ser. Lecture Notes in Computer Science, vol. 8693.\hskip 1em plus 0.5em minus 0.4em\relax Springer, 2014, pp. 740--755.

\bibitem{glue}
A.~Wang, A.~Singh, J.~Michael, F.~Hill, O.~Levy, and S.~R. Bowman, ``{GLUE:} {A} multi-task benchmark and analysis platform for natural language understanding,'' in \emph{7th International Conference on Learning Representations, {ICLR} 2019, New Orleans, LA, USA, May 6-9, 2019}.\hskip 1em plus 0.5em minus 0.4em\relax OpenReview.net, 2019.

\bibitem{wdprune}
F.~Yu, K.~Huang, M.~Wang, Y.~Cheng, W.~Chu, and L.~Cui, ``Width {\&} depth pruning for vision transformers,'' in \emph{Thirty-Sixth {AAAI} Conference on Artificial Intelligence, {AAAI} 2022, Thirty-Fourth Conference on Innovative Applications of Artificial Intelligence, {IAAI} 2022, The Twelveth Symposium on Educational Advances in Artificial Intelligence, {EAAI} 2022 Virtual Event, February 22 - March 1, 2022}.\hskip 1em plus 0.5em minus 0.4em\relax {AAAI} Press, 2022, pp. 3143--3151.

\bibitem{torch_profiler}
\BIBentryALTinterwordspacing
PyTorch, ``Pytorch profiler,'' 2024. [Online]. Available: \url{https://pytorch.org/docs/stable/profiler.html}
\BIBentrySTDinterwordspacing

\bibitem{hvpm}
\BIBentryALTinterwordspacing
M.~S. Inc., ``High voltage power monitor,'' 2024. [Online]. Available: \url{https://www.msoon.com/high-voltage-power-monitor}
\BIBentrySTDinterwordspacing

\bibitem{pytorch}
A.~Paszke, S.~Gross, F.~Massa, A.~Lerer, J.~Bradbury, G.~Chanan, T.~Killeen, Z.~Lin, N.~Gimelshein, L.~Antiga \emph{et~al.}, ``Pytorch: An imperative style, high-performance deep learning library,'' \emph{Advances in neural information processing systems}, vol.~32, 2019.

\bibitem{tf}
TensorFlow, ``Tensorflow lite,'' \url{https://www.tensorflow.org/lite}.

\bibitem{jiang2020mnn}
X.~Jiang, H.~Wang, Y.~Chen, Z.~Wu, L.~Wang, B.~Zou, Y.~Yang, Z.~Cui, Y.~Cai, T.~Yu \emph{et~al.}, ``Mnn: A universal and efficient inference engine,'' \emph{Proceedings of Machine Learning and Systems}, vol.~2, pp. 1--13, 2020.

\bibitem{grpc}
\BIBentryALTinterwordspacing
Google, ``gprc - a rpc library and framework,'' 2024. [Online]. Available: \url{https://grpc.io/}
\BIBentrySTDinterwordspacing

\bibitem{timm}
R.~Wightman, ``Pytorch image models,'' \url{https://github.com/rwightman/pytorch-image-models}, 2024.

\bibitem{wolf-etal-2020-transformers}
T.~Wolf, L.~Debut, V.~Sanh, J.~Chaumond, C.~Delangue, A.~Moi, P.~Cistac, T.~Rault, R.~Louf, M.~Funtowicz, J.~Davison, S.~Shleifer, P.~von Platen, C.~Ma, Y.~Jernite, J.~Plu, C.~Xu, T.~L. Scao, S.~Gugger, M.~Drame, Q.~Lhoest, and A.~M. Rush, ``Transformers: State-of-the-art natural language processing,'' in \emph{Proceedings of the 2020 Conference on Empirical Methods in Natural Language Processing: System Demonstrations}.\hskip 1em plus 0.5em minus 0.4em\relax Online: Association for Computational Linguistics, Oct. 2020, pp. 38--45.

\end{thebibliography}

\newpage

\appendix
\section{Appendix}
In this supplementary material, we introduce how to construct the latency predictor. 
Besides, we show the process of decomposing the large transformer into multiple sub-models and utilize Bayesian optimization to gain the efficient decomposition policy. 
The detailed experimental setup and additional experimental results are presented. 
Table \ref{tab:notation} is a summary of the used notation.

\begin{table}[h]
  \centering
  \caption{Notation.}
  \label{tab:notation}
  \resizebox{\linewidth}{!}{
    \begin{tabular}{c|l} 
      \toprule
      \textbf{Symbol}         & \textbf{Description}                                                                                                                      \\ 
      \hline
      $L$                     & The layer number of the large ViT.                                                                                                        \\
      $d$                     & The embedding dimension of the large ViT.                                                                                                 \\
      $h$                     & The number of attention heads in the large ViT.                                                                                           \\
      $D$                     & The neuron dimension in the MLP block of the large ViT.                                                                                   \\
      $N$                     & The number of heterogeneous edge devices.                                                                                                 \\
      $l_n$                   & The layer number of the sub-model $n$.                                                                                                      \\
      $d_n$                   & The embedding dimension of the sub-model $n$.                                                                                               \\
      $h_n^k$                 & The number of attention heads in the $k$-th layer of the sub-model $n$.                                                                     \\
      $h_n^{1:l_n}$             & \begin{tabular}[c]{@{}l@{}}The set of attention head numbers from the first layer to the $k$-th\\layer of the sub-model $n$.\end{tabular}  \\
      $D_n^k$                 & The neuron dimension in the $k$-th MLP block of the sub-model $n$.                                                                          \\
      $D_n^{1:l_n}$             & \begin{tabular}[c]{@{}l@{}}The neuron dimension from the first MLP block to the $k$-th \\block of the sub-model $n$. \end{tabular}        \\
      $C_n$                   & The architecture specifications of sub-model $n$.                                                                                           \\
      $\mathcal{C}$           & The set of decomposition decisions.                                                                                                       \\
      $\mathcal{N}$           & The set of sub-models.                                                                                                                      \\
      $f(\cdot)$              & The latency predictor.                                                                                                                    \\
      $\mathbf{X}_n$          & The intermediate features of sub-model $n$.                                                                                                 \\
      $r_n$                   & The data transmission rate of the edge device $n$.                                                              \\
      $\mathbf{X}_{agg}$        & The aggregated features from all sub-models.                                                                                                \\
      $\mathbf{W}$            & The weight of linear transformation in the Results Aggregation.                                                                           \\
      $g$                     & The available computing power for the central node $i$.                                                                                   \\
      $t_n^1$                 & The latency of the Phase 1: Backbone Forward of sub-model $n$.                                                                              \\
      $t_n^2$                 & The latency of the Phase 2: Data Transmission of sub-model $n$.                                                                             \\
      $t_n^3$                 & The latency of the Phase 3: Results Aggregation of sub-model $n$.                                                                           \\
      $T$                     & The end-to-end inference latency.                                                                                                         \\
      $\mathcal{D}_{train}$     & The training dataset.                                                                                                                     \\
      $\mathcal{D}_{val}$       & The validation dataset.                                                                                                                   \\
      $\mathcal{L}_n$         & The validation loss of sub-model $n$                                                                                                        \\
      $\mathcal{L}$           & The overall accuracy degradation for collaborative inference.                                                                 \\
      $\delta$                & The balancing hyperparameter.                                                                  \\
      $\Omega_n$              & The maximum amount of computing resources for the edge device $n$.                                                                        \\
      $\Phi_n$                & The maximum amount of memory capacity for the edge device $n$.                                                                            \\
      $\omega(C_n)$           & The floating point operations of the sub-model $n$.                                                                                         \\
      $\phi(C_n)$             & The memory requirements of the sub-model $n$.                                                                                               \\
      $\varOmega$             & The set of all sub-models' floating point operations.                                                                                       \\
      $\varGamma$             & The set of all edge devices' memory requirements.                                                                                         \\
      $\hat{h}$               & The average number of attention heads among all sub-models.                                                                                 \\
      $\hat{D}$               & The average neuron dimension of MLP among all sub-models.                                                                                   \\
      $\varPsi (\mathcal{C})$ & The objective function of latency and accuracy degradation minimization.                                                                      \\
      $\epsilon$              & Gaussian white noise.                                                                                                                     \\
      $M$                     & The number of samples in the training dataset.                                                                                            \\
      $w_i^n$                 & \begin{tabular}[c]{@{}l@{}}The weight of $i$-th sample in the training dataset when \\calibrating the $n$-th sub-model. \end{tabular}          \\
      $W_n$                   & The set of samples' weight when calibrating the $n$-th sub-model.                                                                            \\
      $\mathcal{L}_{CE}$        & The cross entropy loss.                                                                                                                   \\
      $\mathcal{L}_{Bo}^n$         & The calibrating loss of the $n$-th sub-model.                                                                                                \\
      $\varLambda(\cdot)$     & The prediction results of ViT.                                                                                                            \\
      \bottomrule
      \end{tabular}
  }
  
  \end{table}

\begin{table*}
  \centering
  \caption{Comparison between resources on typical computing platforms and deployment demands for representative models.}
  \label{tab:device_model}
  \resizebox{\linewidth}{!}{
      \setlength{\tabcolsep}{3pt}
      \begin{tabular}{c|cccc|c|c|cccc} 
      \toprule
      \multicolumn{5}{c|}{\textbf{Capabilities of Typical Computing Platforms}}                                                                                                    &  & \multicolumn{5}{c}{\begin{tabular}[c]{@{}c@{}}\textbf{Overheads for }\textbf{Deploying~Representative Models~}\end{tabular}}                         \\ 
      \cline{1-5}\cline{7-11}
      \textbf{Devices}        & \textbf{Memory} & \begin{tabular}[c]{@{}c@{}}\textbf{Computation}\\\textbf{Performance}\end{tabular} & \textbf{TDP} & \textbf{Cost} &  & \textbf{Models} & \begin{tabular}[c]{@{}c@{}}\textbf{Required}\\\textbf{Memory}\end{tabular} & \textbf{\textbf{FLOPs}} & \textbf{\#Params.} & \textbf{Storage}  \\ 
      \cline{1-5}\cline{7-11}
      \multicolumn{5}{c|}{\textbf{Commercial GPUs }}                                                                                                                &  & \multicolumn{5}{c}{\textbf{CNNs }}                                                                                                                              \\ 
      \cline{1-5}\cline{7-11}
      V100 SXM3               & 32 GB           & 16.4 TFLOPS                                                                        & 300 W        & \$ 4.4/hr     &  & ResNet50        & 2.2 GB                                                                     & 8.3 G                   & 26 M               & 97.7 MB           \\
      A100 SXM4               & 80 GB           & 19.5 TFLOPS                                                                        & 400 W        & \$ 12.0/hr    &  & ResNet101       & 2.4 GB                                                                     & 15.7 G                  & 45 M               & 170.0 MB          \\
      H100 SXM5               & 80 GB           & 66.9 TFLOPS                                                                        & 700 W        & \$ 20.7/hr    &  & MobileNet V2    & 2.1 GB                                                                     & 0.6 G                   & 4 M                & 13.5 MB           \\ 
      \cline{1-5}\cline{7-11}
      \multicolumn{5}{c|}{\textbf{Edge Devices }}                                                                                                                   &  & \multicolumn{5}{c}{\textbf{Transformers }}                                                                                                                              \\ 
      \cline{1-5}\cline{7-11}
      Raspberry Pi-4B \cite{pi}        & 8 GB            & 13.5 GFLOPS                                                                        & 7.3 W        & \$ 99         &  & Swin-L \cite{swin}         & 3.3 GB                                                                     & 103.9 G                 & 197 M              & 751.0 MB          \\
      NVIDIA Jetson Nano \cite{jetson}     & 4 GB            & 235.8 GFLOPS                                                                       & 10 W         & \$ 60         &  & ViT-L/16 \cite{vit}        & 5.3 GB                                                                     & 123.1 G                 & 304 M              & 1.1 GB            \\
      NVIDIA Jetson TX2 \cite{jetson}      & 8 GB            & 665.6 GFLOPS                                                                       & 15 W         & \$ 249        &  & Flan-T5 Large \cite{flan-t5}         & 4.2 GB                                                                     & 1780 G                 & 751 M              & 2.9 GB          \\
      NVIDIA Jetson Orin Nano \cite{jetson}  & 4 GB            & 640.0 GFLOPS                                                                       & 10 W         & \$ 199        &  & GPT2-XL \cite{gpt2}        & 7.8 GB                                                                     & 3340 G                 & 1560 M              & 5.8 GB            \\
      \bottomrule
      \end{tabular}
    }
\end{table*}

\subsection{Latency Predictor Training}
\label{sec:predictor}
In order to find an optimal policy to minimize collaborative inference latency on heterogeneous edge devices, 
we need to evaluate the hardware latency of decomposed sub-models. There are usually two ways: 
\textit{(1) online measurement:} during the optimization process, directly measure the latency on the target device;
\textit{(2) offline prediction:} utilize a predictor to provide the latency without on-device measurement. 
For the online measurement, a single decomposed sub-model requires hundreds of inferences to get an accurate latency, which lasts for minutes and slows the optimization. 
We apply the offline prediction here because it is fast and accurate. The architecture of sub-models is encoded into a feature vector, and predict its latency instantly with a multi-layer perceptron (MLP). 
The process of training the latency predictor is as follows. 

\textbf{Latency dataset collection.}
Several researchers indicated linear transformation, patch embedding and attention operations are most time-consuming in one transformer inference \cite{efficientformer}. 
The computational cost of these operations mainly depends on the number of transformer layer $l_n$, the embedding dimension $d_n$, the number of attention heads $h_n$ and the neuron dimension $D_n$ of MLP. 
Hence, we evaluate the actual inference latency of transformer models with different architectures on the specific edge device and collect the dataset of \textit{(sub-model architecture, measured latency)} pairs for all edge devices. 

\textbf{Latency predictor fitness.}
The collected latency dataset is utilized to train the latency predictor. The input of latency predictor is a feature vector of sub-model architecture with four elements: the number of transformer layer $l$, the embedding dimension $d$, the average number of attention heads $\bar{h}=\sum_{i=1}^{N}h_i$ and the average neuron dimension of MLP $\bar{D}=\sum_{i=1}^{N}D_i$. 
We utilize a three-layer MLP with 600 hidden dimension and ReLU activation as the predictor $f(l, d, \bar{h}, \bar{D})$ to provide latency because the three-layer MLP is more efficient and accurate. 
Trained with thousands of real latency data points, the predictor yields high prediction accuracy. 
Note that the predicted latency is only used in the optimization process.
Compared with deducing a closed-form latency model for each hardware, the latency predictor method is more general and faster.

\subsection{Decomposer: Bayesian Decomposition}
\label{sec:decomposer}
We utilize the \textit{decomposer} to generate decomposition policies for the large transformer as shown in Lines 1 to 11 of Algorithm 1. 
These policies are evaluated on the \textit{evaluator}. Evaluation results are utilized to update \textit{decomposer} until finding the optimal policy. 
\textit{Decomposer} decomposes the target transformer into multiple sub-models according to the final policy. 
We first present the process of decomposing the target transformer model, and then illustrate how to gain efficient decomposition policies. 

\subsubsection{Transformer Decomposition}
\begin{figure}
  \centering
  \includegraphics[width=0.35\textwidth]{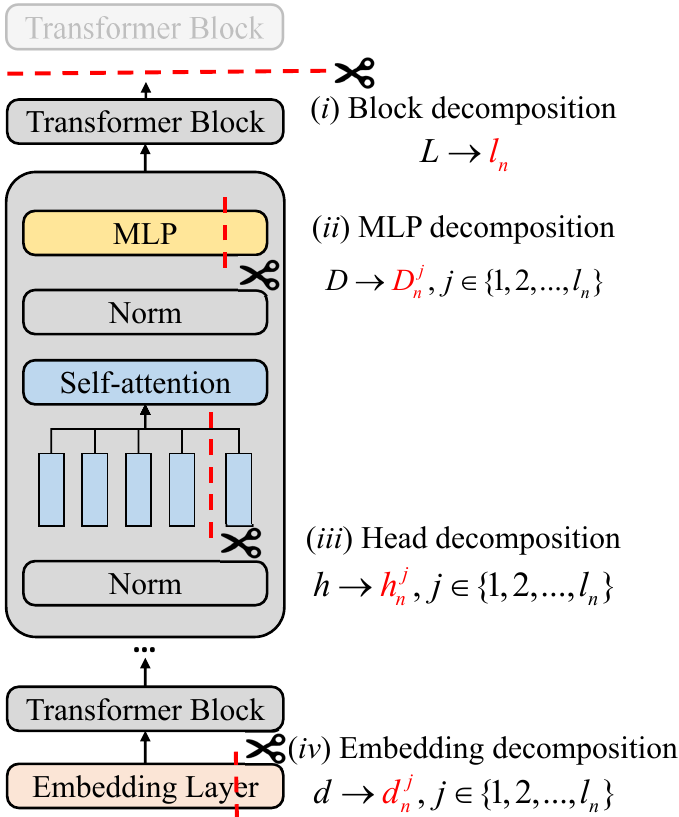}
  \caption{The process of decomposition. The target transformer is decomposed into multiple sub-models according to the number of layers, the number of heads, the MLP dimension and embedding dimension}
  \label{fig:decompose}
\end{figure}

Given $N$ heterogeneous edge devices, the large ViT is decomposed into $N$ distinct sub-ViTs.
As depicted in Figure \ref{fig:decompose}, the decomposition procedure for sub-ViT $n$ encompasses:

\label{sec:tf_decompose}
\begin{itemize}
  \item \textbf{Block decomposition:} The number of layers in the sub-ViT is decreased to $l_n$. 
  \item \textbf{MLP decomposition:} Some unimportant neurons in the MLP block are eliminated. The revised neuron dimension for the $j$-th MLP block is $D_n^j$, where $ j\in\{1,2,...,l_n\}$.
  \item \textbf{Head decomposition:} Some redundant attention heads are removed, resulting in a reduced number of heads in the $j$-th layer to $h_n^j$, where $j\in\{1,2,...,l_n\}$.
  \item \textbf{Embedding decomposition:} The dimension of embedding is diminished to $d_n$.
\end{itemize}

\subsubsection{Optimize Decomposition Policy}
As analyzed in \S III-B3, we treat the objective function in the equation (8) as a black-box-function $\varPsi (\mathcal{C})=\mathcal{L}(\mathcal{C})+\delta T(\mathcal{C})$ to optimize. 
BO \cite{DBLP:journals/pieee/ShahriariSWAF16} is one of the state-of-the-art techniques for black-box optimization. It applies exploration and exploitation to the objective function by sequentially and actively querying the function values of some input instances. 
Specifically, BO uses Gaussian process \cite{williams2006gaussian} as a \textit{prior model} to fit the black-box objective function $\varPsi (\mathcal{C})$ in a closed form. 
It updates the distribution of $\varPsi (\mathcal{C})$ to gain \textit{posterior prediction} by using its likelihood on newly evaluated $(\mathcal{C}, y=\varPsi (\mathcal{C})+\epsilon)$ pairs, 
where $y$ is a noisy observation of $\varPsi (\mathcal{C})$ and is the sum of inference latency and accuracy degradation in our case. 
Then it determines the next decomposition policy $\mathcal{C}$ by minimizing an \textit{acquisition function}, 
which is computed from the updated posterior. The \textit{acquisition function} performs a trade-off between exploration and exploitation in evaluating the candidates of decomposition policy. 
BO repeats the above process until achieving a precise posterior predictive distribution of $\varPsi (\mathcal{C})$. 

\textbf{Prior model.}
We utilize the Gaussian process (GP) as the prior to model the objective function. 
A GP prior is specified by its mean function $\mu(\cdot)=\mathbb{E} [\cdot]$ and covariance function $K(\cdot,\cdot)$. 
The prior of objective function is defined as follows:
\begin{equation}
  \varPsi (\mathcal{C}) \thicksim \text{GP}(\mu(\cdot), K(\cdot,\cdot)),
\end{equation}
We set $\mu(\cdot)=0$ and $K(\cdot,\cdot)$ to be the Matern kernel \cite{DBLP:journals/jmlr/Genton01} with smoothness factor $\nu =1.5$ and length scale $l=1$. 
$K_{\nu}(\cdot)$ and $\Gamma(\cdot)$ are a modified Bessel and gamma function, respectively. The covariance function is defined as follows:
\begin{equation}
  K(\mathcal{C}_i,\mathcal{C}_j)=\frac{\varTheta^{\nu} K_{\nu}\left(\varTheta\right)}{\Gamma(\nu)2^{\nu-1}}, \varTheta=\sqrt{2\nu ||\mathcal{C}_i-\mathcal{C}_j||_2}/l.
\end{equation}

\textbf{Posterior prediction.}
We sample $r$ different decomposition policy $\mathcal{C'}$ satisfying the constraints $\varOmega =\{\Omega_1,...,\Omega_N\}$ and $\varPhi=\{\Phi_1,...,\Phi_N\}$ of all devices, 
and obtain $r$ sets of decomposed sub-models. 
For notation simplicity, vectors composed of $\{\mathcal{C}'\}_{i=1}^{r}$ and $\{\varPsi (\mathcal{C}')\}_{i=1}^{r}$ are denoted as $\mathcal{C}'_{1:r}$ and $\varPsi (\mathcal{C}'_{1:r})$, respectively. 
The GP prior indicates a Gaussian distribution over function values, \ie $\varPsi (\mathcal{C}'_{1:r})|\mathcal{C}'_{1:r} \thicksim \mathbf{N}(0, \mathbf{K})$, where $\mathbf{K}_{i,j}=K(\mathcal{C}'_i,\mathcal{C}'_j), \forall i,j \in [1,r]$. 
We evaluate the objective function denoted by $\hat{y}_i$ as a noisy observation of $\varPsi (\mathcal{C}')$, \ie $\hat{y}_i=\varPsi (\mathcal{C}')+\epsilon$ where Gaussian white noise $\epsilon \thicksim \mathbf{N}(0, \sigma^2)$. 
Given the noisy observations $\hat{y}_{1:r}$, the GP posterior of the black-box function can be updated as follows:
\begin{equation}
  \varPsi (\mathcal{C}'_{1:r})|\mathcal{C}'_{1:r}, \hat{y}_{1:r} \thicksim \mathbf{N}(\hat{y}_{1:r}, \mathbf{K}+\sigma^2\mathbf{I}).
\end{equation}
Given a new decomposition policy $\mathcal{C}$ and $\mathbf{k}_i=K(\mathcal{C},\mathcal{C}'_i)$, the GP posterior can be used to predict the distribution of $\varPsi (\mathcal{C})$: 
\begin{equation}
  \begin{aligned}
    \varPsi (\mathcal{C})|\mathcal{C}'_{1:r},\hat{y}_{1:r} &\thicksim \mathbf{N} (\mu(\mathcal{C}), \sigma^2(\mathcal{C})),\\
    \mu(\mathcal{C}) & \triangleq \mathbf{k}(\mathbf{K}+\sigma^2\mathbf{I})^{-1}\hat{y}_{1:r},\\
    \sigma^2(\mathcal{C}) & \triangleq K(\mathcal{C}, \mathcal{C})-\mathbf{k}^T (\mathbf{K}+\sigma^2\mathbf{I})^{-1}\mathbf{k}.
  \end{aligned}
\end{equation}

\textbf{Acquisition function.}
Given the posterior predictive distribution of $\varPsi(\mathcal{C})$, BO finds the next decomposition policy $\mathcal{C}'_{i+1}$ to evaluate based on an acquisition function $u_i(\mathcal{C})$ defined by the posterior mean $\mu_i(\mathcal{C})$ and standard deviation $\sigma_i(\mathcal{C})$. 
In our algorithm, expected improvement (EI) \cite{jones1998efficient} is utilized as acquisition function. 
For problem (P1), we can determine the next decomposition policy $\mathcal{C}'_{i+1}$ by minimizing EI:
\begin{equation}
  \begin{aligned}
    \mathcal{C}'_{i+1} &= \argmin_{\mathcal{C}} u_i(\mathcal{C}), \\
    u_i(\mathcal{C}) &\triangleq (\varPsi^{*}-\mu)Z ((\varPsi^{*}-\mu)/\sigma) + \sigma H ((\varPsi^{*}-\mu)/\sigma),
  \end{aligned}
\end{equation}
where $Z$ and $H$ represent the probability density function and cumulative distribution function of the standard normal distribution, respectively. 
$\varPsi^{*}$ is the best policy in $\mathcal{C}'_{1:r}$.

\subsection{Detailed Experimental Setup}

\textbf{Prototype.}
We implement our system on a real-world collaborative edge computing testbed that consits of an NVIDIA Jetson Nano with 4 GB memory, an NVIDIA Jetson TX2 with 8 GB memory, an NVIDIA Jetson Orin Nano with 4 GB memory and a switch. 
These devices represent the edge devices with different hardware resources and computing capabilities. The detailed configurations of these devices are shown in the lower left part of Table \ref{tab:device_model}. 
Figure \ref{fig:prototype} shows the implemented hardware platform for CoFormer. These devices are connected via a gigabyte switch TP-LINK TL-SG1008D. For bandwidth control, we use the traffic control tool tc\cite{tc}, 
which is able to limit the bandwidth under the setting value. The maximum bandwidth between devices is fixed at 2 Mb/s.

\textbf{Transformer backbones, tasks and datasets.}
To evaluate the generality of our system on vision and language tasks, we selected three most typical applications in mobile vision and language systems. 
\begin{itemize}
  \item \textbf{Generic-category image recognition.} The vision task aims to recognize the generic category of an image. We select three representative ViT backbones (ViT\cite{vit}, DeiT\cite{deit}, Swin\cite{swin}) and two widely-used classification datasets (CIFAR-100 \cite{cifar} and ImageNet-1K \cite{imagenet_1k}). 
  CIFAR-100 contains 50K training images and 10K testing images, uniformly categorized into 100 classes. 
  ImageNet-1K spans 1000 object classes and contains 1,281,167 training images, 50K validation images and 100K test images. 
  They represent a small and large scale image classification tasks, respectively. 
  \item \textbf{Real-time object detection.} This type of vision tasks aims to detect objects in real-time with fast processing time while maintaining a required level of accuracy. 
  We choose Mask R-CNN \cite{mask-rcnn} with ViT, DeiT and Swin backbone as the detection model in the most widely-used dataset, MS COCO 2017 \cite{coco}, to evaluate. 
  MS COCO 2017 contains 0.33 million images, 1.5 million objects, and 80 categories. 
  \item \textbf{General language understanding.} The language task aims to understand natural language. We select three language transformer backbones (BERT \cite{bert}, GPT2 \cite{gpt2} and Flan-T5 \cite{flan-t5}) and a representative General Language Understanding Evaluation (GLUE) \cite{glue} benchmark to evaluate our method. 
  The GLUE benchmark consists of nine language understanding tasks and cover a diverse range of dataset sizes and degrees of difficulty.
\end{itemize}

\textbf{Baselines.}
We compare CoFormer with state-of-the-art edge deployment methods for transformer models by implementing three types of baselines. 
\begin{itemize}
  \item \textbf{Comparison with large transformer models:} Our method can decompose large transformer models for collaborative inference on multiple edge devices. We select three vision transformer models (ViT-L/16, DeiT-B and Swin-L) and three language transformer models (BERT-Large, GPT2-XL and Flan-T5-Large) to demonstrate the effectiveness of our methods.
  \item \textbf{Comparison with efficient transformer models:}We compare our methods with several representative lightweight transformer models (\ie single-edge solution illustrated in Figure 9 (b)) to show the superiority: T2T-ViT\cite{t2t} incorporates a layer-wise transformation to structuralize the image to tokens. 
  EfficientFormer \cite{efficientformer} introduces a dimension-consistent pure transformer as a design paradigm to reduce latency.
  PoolFormer \cite{poolformer} replaces the original attention module in transformers with the simple spatial pooling operator. 
  MobileViTv2 \cite{mobilevitv2} proposes a separable self-attention method with linear complexity and incorporates CNN structures. 
  \item \textbf{Comparison with collaborative inference methods:} DeViT \cite{devit} is the most relevant baseline of our work, designed for vision transformer inference on multiple homogeneous edge devices. 
  We also compare the ensemble of heterogeneous transformer models including EfficientFormer L1, PoolFormer-S36 and MobileViTv2-100, denoted as Hete-Model-Ens. 
  WDPruning \cite{wdprune} is a weight pruning methods for efficient transformer inference. We adapt WDPruning in different pruning ratios to gain multiple compressed models for collaborative inference.   
\end{itemize}

\textbf{Evaluation metrics.}
We employ the \textit{Top-1 classification accuracy}, \textit{mean average precision (mAP)} and evaluation scores on GLUE benchmark as the overall performance metric of image classification, object detection and language understanding tasks, respectively. 
The \textit{end-to-end latency}, \textit{energy consumption} and \textit{average power} are utilized as the efficiency metrics. 
\textit{(i) Measurements of latency:}
Before measurement, we switch the NVIDIA Jetson devices to the max power mode. 
For example, the max power of NVIDIA Jetson Nano is 10 Watt, and turn off the dynamic voltage and frequency scaling to ensure a steady measurement environment. 
For each transformer model, we run it for once as a warm-up and then record the averaged execution time with 50 runs without break using PyTorch Profiler tool \cite{torch_profiler}. 
The aim of warm-up running is to alleviate the impact of weight loading and PyTorch initiation since we have omitted these overheads. 
We use full precision and batch 1 to execute all operations, which can help robustly execute most models and baselines. 
\textit{(ii) Measurements of energy and power:}
We use a Monsoon High Voltage Power Monitor \cite{hvpm} connected to the edge device to obtain accurate power readings over the course of running a forward pass of each test model 50 times. 
The device settings are the same as that in latency measurements. 
From the power signal reported by the monitor, we derive the average per-inference power and energy consumption by first subtracting background power consumption (i.e., power readings when not running any model) 
and then identifying 50 continuous regions of significantly higher power draw. Each region like that is considered a single inference and we calculate its total energy as the integral over the individual power samples. 
Analogously we also calculate average power consumption by averaging over the same set of samples. After energy and power are calculated for each inference, the final statistics of a model are obtained by again averaging over the 50 identified runs.    
The measurement method follows \cite{DBLP:conf/imc/AlmeidaLMDLL21} for fair comparison.

\textbf{Implementation.}
CoFormer is fully implemented with $~2,000$ LOC in Python in total atop PyTorch \cite{pytorch}. 
Although we use PyTorch for auto-differentiation and computation graph execution, CoFormer is extensible and can work well with other lightweight ML frameworks such as TF-Lite\cite{tf} and MNN\cite{jiang2020mnn}.
The employed transformer models are trained following DeBo algorithm and deployed on all devices in advance. 
The communication module is implemented based on gPRC \cite{grpc}. 
All the compared approaches are run with timm \cite{timm} (vision transformers models) or transformers \cite{wolf-etal-2020-transformers} (language transformer models), for fair comparison.

\subsection{Additional Experimental Results}

\begin{figure*}
  \centering
  \includegraphics[width=0.9\textwidth]{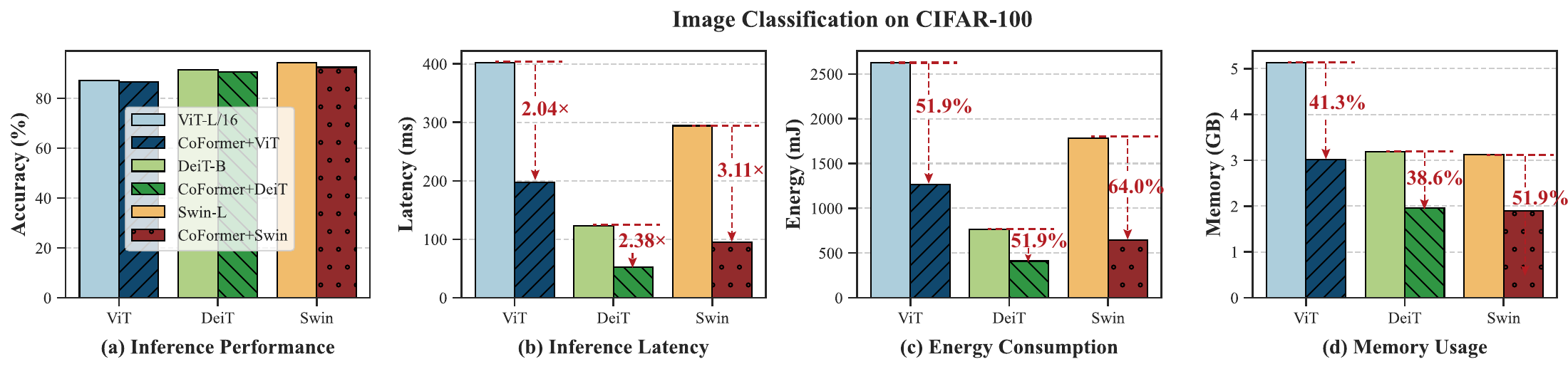}
  \caption{Performance comparison between our methods and large ViTs on CIFAR-100 dataset. Our methods can accelerate inference speed by 2.5 $\times$ and save energy consumption by 54\% on average with slight accuracy sacrifice. }
  \label{fig:re_cifar}
\end{figure*}

\textbf{Performance comparison on CIFAR-100.}
The performance comparison of image classification on the small dataset, CIFAR-100 is presented on Figure \ref{fig:re_cifar}.
The original large transformer models used for the image recognition task include ViT-L/16, DeiT-B, and Swin-L.
Our method can decompose large transformer models into smaller models and achieve efficient collaborative inference while maintaining satisfactory classification accuracy. 
Specifically, our decomposed model, based on Swin-L, can accelerate inference speed by 3.11$\times$, reduce energy consumption by 64.0\% and save memory usage by 51.9\%, on CIFAR-100.
  
\begin{figure}
    \centering
    \includegraphics[width=0.45\textwidth]{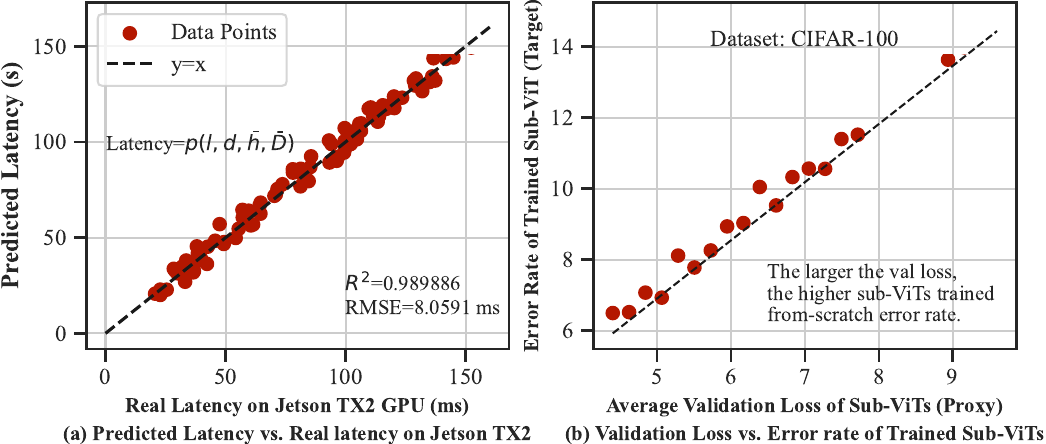}
    \caption{Predicted latency vs. real latency on NVIDIA Jetson TX2 GPU (a) and relationships between validation loss and accuracy degradation of collaborative inference (b).}
    \label{fig:predict}
  \end{figure}

\textbf{Latency predictor and performance proxy.}
We compare predicted latency and real inference latency to validate the performance of the latency predictor. The results of NVIDIA Jetson TX2 on CIFAR-100 dataset are shown in Figure \ref{fig:predict} (top). 
We can observe that predicted latency is very close to real latency with an average prediction error (RMSE) of 8.1 ms. 
As a result, the predictor can yield highly accurate results of inference latency, trained with thousands of real latency data points. 
On the other hand, we utilize the average validation loss of decompose sub-models as the proxy of accuracy degradation. We conduct experiments to show the effectiveness of the proxy in Figure \ref{fig:predict} (bottom). 
The average validation loss of sub-models without training is relative to the error rate of sub-models trained from scratch. 
The large the validation loss of sub-models, the lower the error rate. Therefore, the average validation loss is a good proxy of accuracy degradation for sub-models trained from scratch. 

\vfill

\end{document}